\documentclass[11pt]{article}\pdfoutput=1
\usepackage[nosort]{cite}
\usepackage{amsmath,amsfonts,amssymb}
\usepackage[small,bf,hang]{caption}
\usepackage{slashed}
\usepackage{amsmath}
\usepackage{latexsym,epsfig}
\usepackage{arydshln}
\usepackage[aligntableaux=center,boxsize=0.6em]{ytableau}
\usepackage{genyoungtabtikz}

\usepackage[debug,pageanchor=false]{hyperref}
\hypersetup{colorlinks=true,linktocpage,breaklinks,
            urlcolor=blue,
            linkcolor=red,
            citecolor=blue
            }

\def\hybrid{
        \topmargin -20pt
        \oddsidemargin 0pt
        \headheight 0pt \headsep 0pt
        \textwidth 6.25in 
        \textheight 9.5in 
        \marginparwidth .875in
        \parskip 5pt plus 1pt \jot = 1.5ex}

\hybrid

 \def\b{\beta}  \def\d{\delta} 
\def\ve{\varepsilon}  \def\h{\eta} 
   \def\l{\lambda} \def\m{\mu}
\def\n{\nu} \def\x{\xi}   \def\r{\rho}
 \def\s{\sigma}   
  \def\y{\psi} \def\w{\omega}

   \def\L{\Lambda} 
   
 \def\S{\Sigma}

\def\ba{\bar{a}}\def\be{\bar{e}}

\def\bA{\bar{A}}\def\bB{\bar{B}}\def\bC{\bar{C}}\def\bD{\bar{D}}\def\bE{\bar{E}}
\def\bK{\bar{K}}\def\bL{\bar{L}}\def\bM{\bar{M}}\def\bN{\bar{N}}\def\bP{\bar{P}}\def\bQ{\bar{Q}}
\def\bR{\bar{R}}

\def\fr{\frac}  \def\dt{\partial}

\def\ph{\phantom}
\def\mc{\mathcal}

\def\mL{\mathcal{L}}
\def\mD{\mathcal{D}}

\def\cM{\mathcal{M}}

\def\TT{\mathbb{T}}
\def\SS{\mathbb{S}}

 \csname
@addtoreset\endcsname{equation}{section}

\def\moth{\mathsurround=0pt}
\newdimen\zo \zo=0pt

\def\tick{\leaders\hrule height 0.5ex depth 0pt \hskip 0.5pt}
\def\upboxfill{$\moth \setbox\zo\hbox{\tick}%
  \hskip 3pt\hbox to 0pt{$\tick$\hss}\hrulefill \hbox to 7.5pt{$\tick$\hss}$}

\def\dtick{\leaders\hrule height .34pt depth 0.5ex \hskip 0.5pt}
\def\downboxfill{$\moth \setbox\zo\hbox{\dtick}%
  \hskip 2pt\hbox to 0pt{$\dtick$\hss}\hrulefill \hbox to 2pt{$\dtick$\hss}$}

\def\bec{\begin{center}}
\def\ec{\end{center}}

\def\b{\beta}

\def\ve{\varepsilon}

\def\y{\eta}

\def\cB{{\cal B}}

\def\cA{{\cal A}}

\def\cM{{\cal M}}

\def\cA{{\cal A}}

\def\be{\begin{equation}}
\def\ee{\end{equation}}
\def\bea{\begin{eqnarray}}
\def\eea{\end{eqnarray}}
\def\ba{\begin{array}}
\def\ea{\end{array}}

\begin{document}

\begin{titlepage}

\vfill
\begin{flushright}
\end{flushright}

\vfill

\begin{center}

\mathversion{bold}

   \baselineskip=16pt
   {\LARGE \bf  ${\rm O}(d\!+\!1,d\!+\!1)$ enhanced double field theory}
   \vskip 2cm

\mathversion{normal}

{\large\bf {Olaf Hohm${\,}^1$, Edvard T. Musaev${\,}^{2,3}$, Henning Samtleben${\,}^4$}}
\vskip .6cm
{\it ${}^1$
Simons Center for Geometry and Physics, Stony Brook University,\\
Stony Brook, NY 11794-3636, USA}\\
ohohm@scgp.stonybrook.edu
\vskip .2cm

{\it ${}^2$
Max-Planck-Institut f\"ur Gravitationsphysik (Albert-Einstein-Institut)\\
Am M\"uhlenberg 1, DE-14476 Potsdam, Germany}\\
 edvard.musaev@aei.mpg.de
\vskip .2cm

{\it $^{3}$Kazan Federal University, Institute of Physics, General Relativity Department, \\
                          Kremlevskaya 16a, 420111, Kazan, Russia}
\vskip .2cm

{\it ${}^4$ Univ Lyon, ENS de Lyon, Univ Claude Bernard Lyon 1, CNRS, Laboratoire de Physique, F-69342 Lyon, France} \\
{henning.samtleben@ens-lyon.fr}

\end{center}

\vfill 
\begin{center} 
\textbf{Abstract}

\end{center} 
\begin{quote}

Double field theory yields a formulation of the low-energy effective action 
of bosonic string theory and half-maximal supergravities that 
is covariant under the T-duality group O$(d,d)$ emerging on a torus $T^d$.
Upon 
reduction to three spacetime dimensions and dualisation
of vector fields into scalars, the symmetry group is 
enhanced to O$(d+1,d+1)$.  
We construct an enhanced 
double field theory with internal coordinates in the adjoint representation  
of O$(d+1,d+1)$. 
Its section constraints admit two inequivalent solutions,
encoding in particular the embedding of $D=6$ chiral and non-chiral theories, respectively.
As an application we 
define consistent 
generalized Scherk-Schwarz reductions using  a novel notion 
of generalized parallelization. 
This allows us to prove the consistency of the truncations of 
$D=6$, ${\cal N}=(1,1)$ and $D=6$, ${\cal N}=(2,0)$ supergravity on AdS$_3\times \SS^3$\,.

\end{quote} 
\vfill
\setcounter{footnote}{0}
\end{titlepage}

\clearpage
\setcounter{page}{2}

\tableofcontents

\newpage

\section{Introduction}

The T-duality property of closed string theory implies the emergence of an  
${\rm O}(d,d,\mathbb{R})$ symmetry upon reduction of the low-energy 
effective actions on a torus $T^d$. 
This holds for bosonic string theory but also for the maximal and half-maximal supergravities 
in $D=10$ and their lower-dimensional descendants. 
The O$(d,d)$ invariance is a `hidden' symmetry from the point of view of conventional 
(super-)gravity in that it cannot be explained in terms of the symmetries present before 
compactification. Double field theory (DFT) is the framework that makes O$(d,d)$ manifest 
before reduction by working on a suitably generalized, 
doubled space \cite{Siegel:1993th,Hull:2009mi,Hohm:2010jy,Hohm:2010xe}. 
DFT can be defined for 
the universal NS sector consisting of metric, $b$-field and dilaton, including 
bosonic string theory in $D=26$ and minimal supergravity in $D=10$, but also 
for type II string theory \cite{Hohm:2011zr,Hohm:2011dv}.

The group O$(d,d)$ is the universal duality symmetry arising for toroidal compactification 
of any string theory, but for special theories or backgrounds this symmetry may be enhanced further. 
For instance, for half-maximal supergravity coupled to $n$ vector multiplets 
(or heterotic string theory with $n=16$) 
the symmetry is enhanced to O$(d,d+n)$, for which there is a 
DFT formulation \cite{Siegel:1993th,Siegel:1993bj,Hohm:2011ex}. 
Moreover, compactifications of half-maximal supergravity to $D=4$ 
also exhibit an SL$(2)$ duality, for which a DFT formulation has been obtained recently \cite{Ciceri:2016hup}. 
The case of interest for the present paper is the compactification to three spacetime dimensions. 
In this case, $D=10$ supergravity yields  an O$(7,7)$ symmetry that,  
however, is enhanced to O$(8,8)$ for half-maximal and 
to E$_{8(8)}$ for maximal supergravity. Similarly, heterotic string theory exhibits 
an enhanced O$(8,24)$ duality \cite{Sen:1994wr}, while the T-duality group of $D=26$  
bosonic string theory on $T^{23}$ is enhanced to O$(24,24)$. More generally, a string theory 
compactified on $T^d$ to three spacetime dimensions exhibits an O$(d+1,d+1)$ symmetry. 
This comes about because vector fields in three dimensions can be dualized into scalars 
which join the universal scalars to combine into a larger coset model \cite{Breitenlohner:1987dg,Cremmer:1999du}. 

Our goal in this paper is to define an `enhanced double field theory' that makes  the larger duality group 
O$(d+1,d+1)$ manifest before compactification by working on a suitable extended internal space. 
More generally, we will define the theory for any pseudo-orthogonal group O$(p,q)$. 
In this we closely follow the construction of the maximal E$_{8(8)}$ exceptional field theory \cite{Hohm:2014fxa} and 
the SL$(2,\mathbb{R})$ covariant formulation of $D=4$ Einstein gravity \cite{Hohm:2013jma}. 
Concretely, we generalize the formulation of \cite{Hohm:2013nja} 
to an 
enhanced double field theory, with external and (extended) 
internal coordinates,
but the internal coordinates now live in the adjoint representation of O$(p,q)$.\footnote{
Since the coordinates are split into external and internal, with the latter not only being doubled but embedded into the adjoint representation of O$(p,q)$,  
this theory could also be referred to as an exceptional field theory in the sense of \cite{Hohm:2014fxa}. 
We thank the referee for pointing this out. 
} 
The coordinates thus read $Y^{\cal M}=Y^{[MN]}$ with fundamental indices $M,N,\ldots =1,\ldots, p+q$, subject to 
section constraints that generalize the level-matching constraint of DFT. 
A novel feature of this theory compared to the original DFT is that the section constraint 
has inequivalent solutions. As a consequence, we can embed in particular both the chiral and non-chiral theories
in $D=6$.

One of the conceptually most intriguing aspects of double and exceptional field theories 
with three external dimensions is that they require the inclusion of `dual graviton' degrees of freedom. 
Indeed, in dimensional reduction the three-dimensional vector fields need to be dualized into scalars 
in order to realize the enhanced symmetry, and these vectors 
include the Kaluza-Klein vector fields originating from the higher-dimensional metric. 
Thus, their duals would be part of a higher-dimensional dual graviton, whose existence  within   
a more or less conventional field theory is excluded by strong no-go theorems \cite{Bekaert:2004dz}.
This is reflected in the observation that the generalized Lie derivatives supposed to unify the internal 
diffeomorphisms and tensor gauge transformations do not define 
a consistent gauge algebra for duality groups associated to three dimensions such as O$(8,8)$ \cite{Strickland-Constable:2013xta}. 
Within exceptional field theory this obstacle shows up in the gauge transformations 
of the tensor hierarchy in any dimension $n$, among the gauge symmetries associated to 
the $(n-2)$-forms \cite{Hohm:2013pua,Hohm:2013vpa,Hohm:2013uia,Godazgar:2014nqa}. 
Nevertheless, consistent double and exceptional field theories can be defined 
upon including an additional gauge symmetry (subject to somewhat unusual constraints) and 
its associated gauge potential. Three external dimensions are special because the need for 
additional gauge symmetries is apparent already at the level of the `scalar' fields, and the additional 
gauge potential features among the `vectors' participating in the gauging and the needed 
Chern-Simons action.

Concretely, the internal (generalized) diffeomorphisms  
parameterized by $\Lambda^{\cal M}$ have to be augmented by new 
gauge symmetries with parameters $\Sigma_{\cal M}$ that are subject to `extended sections constraints' 
requiring that they behave like a derivative in that, e.g., $\Sigma^{\cal M}\partial_{\cal M}=0$. Nevertheless, 
this additional gauge parameter \textit{cannot} be reduced to the derivative of a (singlet) gauge 
parameter, nor can the associated gauge vector be eliminated in terms of (derivatives of) the other gauge fields. 
In the present paper we will confirm that precisely the same 
construction applies to enhanced DFT with duality group O$(p,q)$. 
Moreover, we use the opportunity to clarify the properties of these enhanced gauge symmetries 
by showing that on the space of `doubled' gauge parameters $\Upsilon\equiv(\Lambda^{\cal M},\Sigma_{\cal M})$
one has a generalized Dorfman product that shares all properties familiar from, say, DFT. 
In particular, we will show that the Chern-Simons action can be naturally defined in terms 
of a similarly `doubled' gauge field $ \mathfrak{A}_{\mu}  \equiv ({\cal A}_{\mu}{}^{\cal M}, {\cal B}_{\mu\, {\cal M}})$.

As one of our main applications we will use the O$(p,q)$ DFT to define consistent generalized 
Scherk-Schwarz compactifications as in \cite{Lee:2014mla,Hohm:2014qga}, 
employing a novel notion of generalized parallelization. 
For a generalized Scherk-Schwarz reduction, the compactification data are entirely encoded in a 
group matrix (`twist matrix') $U^{\cal N}{}_{\bar{\cal M}}$ and a singlet $\rho$, both depending only on the internal 
coordinates $Y^{\cal M}$.  For duality group O$(p,q)$ the twist matrix can be decomposed into fundamental 
matrices $U{}^{N}{}_{\bar{M}}$, and we define a `doubled' twist matrix as for the gauge parameters and 
gauge fields: 
 \be\label{frakUintro}
  \mathfrak{U}_{\bar{M}\bar{N}} 
  \ \equiv \ \big(\rho^{-1} U{}^{K}{}_{[\bar{M}} U{}^{L}{}_{\bar{N}]}\;, \;-\frac{1}{4}\,\rho^{-1}(\partial_{KL} U^{P}{}_{\bar{M}})   
  U_{P\bar{N}}\big)\;. 
 \ee
Although at the level of elementary gauge fields and parameters the additional (covariantly constrained) 
components cannot be eliminated in terms of (derivatives of) the other fields, for the Scherk-Schwarz 
ansatz the corresponding component $\mathfrak{U}_{\bar{M}\bar{N}\,KL}$ can be written in terms 
of derivatives of the twist matrix. Note that with its indices being carried by a derivative, the above form 
is manifestly consistent with the constraint. 
We will show that a twist matrix gives rise to a consistent compactification provided the doubled tensor 
(\ref{frakUintro}) satisfies the following algebra with respect to the (generalized) Dorfman product $\circ$:  
 \be\label{TwistIntro}
  \mathfrak{U}_{\bar{M}\bar{N}} \ \circ \ \mathfrak{U}_{\bar{K}\bar{L}} \ = \
   -X_{\bar{M}\bar{N},\bar{K}\bar{L}}{}^{\bar{P}\bar{Q}}\, \mathfrak{U}_{\bar{P}\bar{Q}}\;, 
 \ee  
where the $X$ are constant  and define the embedding tensor of gauged supergravity.  
For the `geometric component' this relation encodes the familiar Lie algebra of Killing vector fields. 
The above defines a notion of generalized parallelizability. 
Writing the  compactification ansatz in terms of the twist matrix, for instance 
for the `doubled' gauge vector as $\mathfrak{A}_{\mu}(x,Y) =  \mathfrak{U}_{\bar{M}\bar{N}}(Y) 
A_{\mu}{}^{\bar{M}\bar{N}}(x)$, we will show that the $U$-matrices and hence the $Y$-dependence 
factors out homogeneously, thus proving consistency of the compactification. 
We will thereby prove the consistency of a large class of compactifications to three dimensions, 
including the truncations of 
$D=6$, ${\cal N}=(1,1)$ and $D=6$, ${\cal N}=(2,0)$ supergravity on AdS$_3\times \SS^3$.

This paper is organized as follows. In sec.~2 we introduce the 
${\rm O}(p,q)$ generalized diffeomorphisms, the generalized Dorfman product 
and the associated gauge vectors. Based on this, we construct in sec.~3 the complete ${\rm O}(p,q)$
enhanced DFT, and discuss its relation, upon solving the section constraint, to (super-)gravity 
theories in various dimensions. In sec.~4 we discuss generalized Scherk-Schwarz compactifications 
in terms of generalized parallelizability and analyze the `twist equations' 
(\ref{TwistIntro}). These results are then applied in sec.~5 in order to establish the consistency 
of various Kaluza-Klein truncations to three dimensions. We conclude in sec.~6 with a general outlook 
on further applications and generalizations.  Appendix A collects some ${\rm O}(p,q)$ identities, 
and in appendix B we give for completeness the details of the generalized Dorfman product 
for (doubled) vectors in the case of E$_{8(8)}$.

\section{\texorpdfstring{${\rm O}(p,q)$}{O(p,q)} generalized diffeomorphisms and tensor hierarchy}

In this section we introduce the ${\rm O}(p,q)$ covariant generalized Lie derivatives that define 
generalized diffeomorphisms. Their structure follows \cite{Hohm:2013jma,Hohm:2014fxa}
and is conceptually different from theories with external dimension 
$n\geq 4$: they are defined with respect to a pair of gauge parameters, one of which is subject to further 
constraints. We clarify their algebraic structure by defining a generalized Dorfman product on the space 
of such pairs. This significantly simplifies the subsequent constructions, including the tensor hierarchy 
and the definition of the Chern-Simons action.

\subsection{Generalized diffeomorphisms}\label{firstsubsection}

We begin by spelling out our conventions for the group ${\rm O}(p,q)$. Its fundamental 
representations is indicated by indices $M,N,\ldots = 1,\ldots, p+q$. Hence, 
objects living in the adjoint representation, like the coordinates $Y^{\cal M}$, are labelled by 
index pairs: 
 \be\label{adjointcoord}
  Y^{\cal M} \ \equiv \ Y^{[MN]} \ \equiv \ Y^{MN}\;.  
 \ee
The structure constants are given by  
\bea\label{IVten1}
f^{MN,KL}{}_{PQ}&=&8\,\delta_{[P}{}^{[M} \eta^{N][K} \delta_{Q]}{}^{L]}
\;,
\eea
with the ${\rm O}(p,q)$ invariant metric $\eta_{MN}$,
which we use in the following to raise and lower indices.  
In addition, for ${\rm O}(p,q)$ we use two more invariant tensors: 
\bea\label{IVten2}
s^{PQ,MN}{}_{KL}&=&8\,\delta_{(K}{}^{[P} \eta^{Q][M} \delta_{L)}{}^{N]}
\;,
\eea
which is symmetric under exchange of $[PQ]$ with $[MN]$, and 
\bea\label{IVten3}
\mathbb{A}^{PQRS}{}_{KLMN} &\equiv& \delta_{KLMN}{}^{PQRS}~\equiv~
\frac1{24}\left(
\delta_K{}^P\delta_L{}^Q\delta_M{}^R\delta_N{}^S \pm \; {\rm permutations}\;\right)
\;, 
\eea
which is totally antisymmetric in the lower and upper sets of indices. 
  
We can now define section constraints for the derivatives $\partial_{{\cal M}}=\partial_{MN}$ 
dual to the adjoint coordinates  (\ref{adjointcoord}) in analogy to other double and exceptional field 
theories. In terms of the above defined ${\rm O}(p,q)$ tensors, we impose 
\bea
s^{MNKL}{}_{PQ}\,\partial_{MN}\otimes\partial_{KL} &=&0
\qquad
f^{MNKL}{}_{PQ}\,\partial_{MN}\otimes\partial_{KL} ~=~0
\;,\nonumber\\
{\mathbb A}^{MNKL}{}_{PQRS}\,\partial_{MN}\otimes\partial_{KL} &=&0
\;,\qquad
\eta^{MK}\eta^{NL}\, \partial_{MN}\otimes\partial_{KL} ~=~0
\;.
\eea
Writing out the invariant tensors in terms of $\eta$ and Kronecker deltas it is easy to see that 
the section constraints are equivalent to 
\bea
\partial_{[MN}\otimes \partial_{KL]} \ = \  0 ~=~ \eta^{NK}\,\partial_{MN}\otimes \partial_{KL}
\;, 
\label{section}
\eea
which is the form we will use from now on. We recall that as for higher-dimensional DFTs and ExFTs 
these constraints are meant to hold for arbitrary functions and their products, so that for instance for 
fields $A, B$ we impose $\partial_{[MN}A\, \partial_{KL]}B=0$ and $\partial_{M}{}^{K}A\, \partial_{NK}B=0$. 
The constraints simplify when the second-order differential operator acts on a 
single object $A$ as follows 
 \be\label{simplifiedSec}
  0 \ = \  \partial_{M[N}\partial_{PQ]}A \quad \Rightarrow \quad \partial_{MN}\partial_{PQ}A \ = \ 
  -2\,\partial_{M[P}\partial_{Q]N}A\;. 
 \ee
This can be verified by using that partial derivatives commute, 
$\partial_{MN}\partial_{KL}A=\partial_{KL}\partial_{MN}A$.

 \medskip

We now turn to the definition of generalized Lie derivatives acting on arbitrary 
${\rm O}(p,q)$ tensors. For a tensor $V^{MN}$ in the adjoint representation 
it is defined as 
\bea
{\cal L}_{(\Lambda,\Sigma)} V^{MN} &\equiv&
\Lambda^{KL}\partial_{KL}\,V^{MN}
+2\,(p+q-2)\,\mathbb{P}^{PQ}{}_{RS}{}^{MN}{}_{KL}\,\partial_{PQ}\Lambda^{RS}\,V^{KL}
+\lambda\, \partial_{KL}\Lambda^{KL}\, V^{MN}
\nonumber\\
&& -\Sigma_{PQ}\,f^{PQ,MN}{}_{KL}\,V^{KL}
\;,
\label{genLie0}
\eea
where $\mathbb{P}^{\cal M}{}_{\cal N}{}^{\cal K}{}_{\cal L}$ is the projector to the adjoint representation, 
explicitly given in (\ref{ProjAdj}),
and we have also allowed for an arbitrary density weight $\lambda$.
While these terms capture the generic structure of generalized diffeomorphisms~\cite{Coimbra:2011ky,Berman:2012vc}
the last term describes a  local adjoint ${\rm O}(p,q)$ transformation with parameter $\Sigma_{MN}$  
which, subject to constraints, will be seen momentarily to be necessary for consistency. 
Its presence is typical for ExFTs with three external dimensions~\cite{Hohm:2013jma,Hohm:2014fxa}.
The projector $\mathbb{P}$ can be written in terms of the above invariant ${\rm O}(p,q)$ tensors, 
so that one obtains for the generalized Lie derivative 
\begin{equation}
\begin{aligned}
&{\cal L}_{(\Lambda,\Sigma)} V^{MN} =\\
&=
\Lambda^{KL}\partial_{KL}\,V^{MN}
-V^{KL}\,\partial_{KL}\Lambda^{MN}
+(\lambda-1)\,\partial_{PQ}\Lambda^{PQ}\,V^{MN}
\\
&+\Big(
6\,{\mathbb A}^{PQMN}{}_{RSKL}{} 
+\frac{1}{16}\,s^{PQ,MN}{}_{UV} s_{RS,KL}{}^{UV} 
+\frac{1}{16}\,f^{PQ,MN}{}_{UV} f_{RS,KL}{}^{UV}\Big)
\partial_{PQ}\Lambda^{RS}\,V^{KL}
\\
&-\Sigma_{PQ}\,f^{PQ,MN}{}_{KL}\,V^{KL}
\;. 
\label{genLie}
\end{aligned}
\end{equation}
Let us emphasize that 
in the following we will always refer to $\lambda$ as the density weight of a field, as opposed to the 
`effective weight' $(\lambda-1)$ emerging  in the first line of (\ref{genLie}). 

In the following we will show that the generalized Lie derivatives form a closed algebra,  
which in turn fixes the coefficient $2\,(p+q-2)$ in front of the projector in (\ref{genLie0}). 
More precisely, the ${\cal L}_{\Lambda}$ for $\Sigma=0$ do not close separately, but closure 
follows upon including a `covariantly constrained' parameter $\Sigma_{MN}$ satisfying 
the same constraints as the derivatives $\partial_{MN}$: 
\bea
\Sigma_{[MN}\otimes \partial_{KL]} \ = \  0  \ = \  \eta^{NK}\,\Sigma_{MN}\otimes \partial_{KL}
\;,\qquad \mbox{etc.}\;.
\label{sectionSigma}
\eea
 Indeed, defining the gauge variations of a generic tensor field $V$ by the generalized 
 Lie derivative, $\delta_{\Lambda,\Sigma}V\equiv {\cal L}_{(\Lambda, \Sigma)}V$, and 
 provided the section conditions (\ref{section}) are satisfied, one finds for the gauge algebra 
 \bea\label{closure}
  \big[\,\delta_{(\Lambda_{1},\Sigma_1)},\,\delta_{(\Lambda_{2},\Sigma_2)}\,\big] &=& 
  \delta_{[(\Lambda_{2},\Sigma_2),(\Lambda_{1},\Sigma_1)]}\;, 
 \qquad
 [(\Lambda_{2},\Sigma_2),(\Lambda_{1},\Sigma_1)] ~\equiv~ (\Lambda_{12},\Sigma_{12})
 \;,\quad
 \eea
with the effective parameters
\bea
\Lambda_{12}{}^{MN} &=&
2\,\Lambda_{[2}{}^{KL}\partial_{KL} \Lambda_{1]}{}^{MN}
-6\,{\mathbb A}{}^{MNKL}{}_{PQRS}\,\Lambda_{[2}{}^{PQ}\,\partial_{KL} \Lambda_{1]}{}^{RS}
\nonumber\\
&&{}
-\frac1{16}\left(s_{PQRS}{}^{UV} s^{MNKL}{}_{UV}+
f_{PQRS}{}^{UV} f^{MNKL}{}_{UV}\right)
\Lambda_{[2}{}^{PQ}\,\partial_{KL} \Lambda_{1]}{}^{RS}
\;,\nonumber\\[2ex]
\Sigma_{12\,MN}&=&
-2\,\Sigma_{[2\,|MN} \partial_{KL}\Lambda_{1]}{}^{KL}
+2\,\Lambda_{[2}{}^{KL} \partial_{KL}\Sigma_{1]MN}
-2\,\Sigma_{[2\,|KL} \partial_{MN}\Lambda_{1]}{}^{KL}
\nonumber\\
&&{}
-\frac18\,f_{RSKL}{}^{PQ}\,\Lambda_{[2}{}^{RS} 
\partial_{MN}\partial_{PQ} \Lambda_{1]}{}^{KL}
\;. 
\label{LS12}
\eea

In order to prove the above closure result it is convenient (and sufficient) to work with the 
Lie derivative acting on an object in the fundamental representation of ${\rm O}(p,q)$, 
i.e., a vector $V^M$, for which we write 
 \be\label{genLieFUND}
  {\cal L}_{(\Lambda, \Sigma)}V^M \ = \ \Lambda^{KL}\partial_{KL}V^M
  +K^{M}{}_N(\Lambda,\Sigma)\,V^N + \lambda \, \partial_{KL}\Lambda^{KL}\,V^M\;, 
 \ee 
where we defined 
 \be\label{KDEFF}
 K^{MN}(\Lambda,\Sigma)  \ = \ 4\,\big(\partial^{[M}{}_{K}\Lambda^{N]K} + \Sigma^{MN}\big)
  \;.
 \ee 
The action of the generalized Lie derivative on a tensor  with an arbitrary number of fundamental 
${\rm O}(p,q)$ indices is then defined straightforwardly, with a $K$ term for each index.  
In particular, one may verify that this definition reproduces the above form of the generalized Lie 
derivative acting on an adjoint vector $V^{MN}$. 

Closure of the gauge transformations given by the generalized Lie derivatives (\ref{genLieFUND}) 
can now be proved by a direct computation. Specifically, one may quickly verify that closure is 
equivalent to the following condition on $K$: 
 \be\label{KClosureREL}
  K^{M}{}_{N}(\Lambda_{12},\Sigma_{12}) \ = \ \Lambda_2^{KL}\partial_{KL}K^{M}{}_{N}(\Lambda_1,\Sigma_1)
  +K^{M}{}_K(\Lambda_2,\Sigma_2)K^{K}{}_N(\Lambda_1,\Sigma_1) - (1\leftrightarrow 2)\;, 
 \ee 
where $\Lambda_{12}$ and $\Sigma_{12}$, given in (\ref{LS12}), can be simplified 
by writing out the invariant tensors in terms of (\ref{IVten1})--(\ref{IVten3}): 
 \be\label{simplifiedGaugeAlgebra}
  \begin{split}
   \Lambda_{12}^{MN} \ = \ &\, 2\,\Lambda_{[2}{}^{KL}\partial_{KL}\Lambda_{1]}{}^{MN}
   -6\,\Lambda_{[2}{}^{[MN}\partial_{KL}\Lambda_{1]}{}^{KL]}
   -4\,\Lambda_{[2K}{}^{[M}\partial_L{}^{N]}\Lambda_{1]}{}^{KL} \;, \\
   \Sigma_{12MN} \ = \ &\, -2\,\Sigma_{[2\,|MN} \partial_{KL}\Lambda_{1]}{}^{KL}
+2\,\Lambda_{[2}{}^{KL} \partial_{KL}\Sigma_{1]MN}
-2\,\Sigma_{[2\,|KL} \partial_{MN}\Lambda_{1]}{}^{KL}\\
&\,- \Lambda_{[2}{}^{P}{}_{K}\partial_{MN}\partial_{PL}\Lambda_{1]}{}^{KL}\;. 
  \end{split}
  \ee   
As a help for the reader and an illustration of the use of the section constraints (\ref{section}) 
and the analogous constraints (\ref{sectionSigma}) on $\Sigma$, we display the relevant terms  
involving $\Sigma$. It is easy to see that, as a consequence of the constraints, they are linear in $\Sigma$
and vanish  by use of the first constraint in (\ref{sectionSigma}) in the form 
 \be\label{Sigmasection}
  0 \ \equiv \ 6\,\Sigma_{[MK}\partial_{NP]} \ = \ 2\,\Sigma_{M[K}\,\partial_{|N|P]} 
  +2\,\Sigma_{N[P}\,\partial_{|M|K]} -\Sigma_{MN}\partial_{KP}-\Sigma_{KP}\partial_{MN}\;.
 \ee 
 
We will next  discuss the transformation rules for partial derivatives of tensor fields. 
  Let us compute the  variation of the partial derivative of a vector of weight $\lambda$: 
  \be\label{noncovVARPARTIALV}
  \begin{split}
   \delta_{\Lambda,\Sigma}
   (\partial_{MN}V_K) \ = \ &\, \partial_{MN}\big(\Lambda^{PQ}\partial_{PQ}V_K
   +K_K{}^{P}(\Lambda, \Sigma) V_P+\lambda\,  \partial_{PQ}\Lambda^{PQ} \, V_K \big)\\
    \ = \ &\, \Lambda^{PQ}\partial_{PQ}\partial_{MN}V_K
   + \partial_{MN}\Lambda^{PQ}\,\partial_{PQ}V_K  + K_K{}^{P}\partial_{MN}V_P \\
   &\,  +\lambda\,  \partial_{PQ}\Lambda^{PQ} \, \partial_{MN} V_K
   +\partial_{MN}K_K{}^{P}V_P
   +\lambda\,   \partial_{MN}\partial_{PQ}\Lambda^{PQ} \, V_K\;. 
  \end{split} 
  \ee 
In order to compare this with the expression for a generalized Lie derivative, we use the section constraint as in (\ref{Sigmasection}), which yields 
 \be
  \partial_{MN}\Lambda^{PQ}\partial_{PQ}V_K \ = \ 2\, K_{[M}{}^{P}\partial_{|P|N]}V_K
  -\partial_{PQ}\Lambda^{PQ}\partial_{MN}V_K\;. 
 \ee 
Thus, using this in (\ref{noncovVARPARTIALV}), we have shown 
 \be\label{noncovVAR}
   \delta_{\Lambda,\Sigma}(\partial_{MN}V_K) \ = \ {\cal L}_{(\Lambda,\Sigma)}^{[\lambda-1]}(\partial_{MN}V_K)
   +\partial_{MN}K_{K}{}^{P} V_P
    +\lambda\,   \partial_{MN}\partial_{PQ}\Lambda^{PQ} \, V_K\;, 
 \ee  
where the notation in the first term indicates that the generalized Lie derivative acts now with weight $(\lambda-1)$. 
[We will use similar notations in the following whenever it is convenient.]
The additional terms involving second derivatives of the gauge parameter 
are referred to as non-covariant variations. The non-covariant variations for the 
(first) partial derivatives of arbitrary tensors take the analogous form, with a $\partial K$ term for each 
index and one term proportional to $\lambda$ involving $\partial_{MN}(\partial_{PQ}\Lambda^{PQ})$
(which, of course, vanishes for zero density weight).

\medskip

We close this subsection by discussing  trivial gauge parameters or gauge symmetries of gauge symmetries, 
i.e., choices of $(\Lambda,\Sigma)$
whose generalized Lie derivative (\ref{genLie0}) gives zero on all fields as a consequence of the constraints. 
The simplest example is 
 \be\label{Trivial1}
  \Lambda^{MN} \ = \ \partial_{KL}\chi^{[MNKL]}\;, 
 \ee 
with $\Sigma_{MN}=0$.  
Indeed, the transport term vanishes by the section constraint, and $K^{MN}=0$ as a consequence of the section constraint 
in the form  (\ref{simplifiedSec}). There are more subtle trivial gauge parameters, involving both $\Lambda$ and 
$\Sigma$, parameterized by an arbitrary $\chi^{KL}$: 
 \be\label{TRIVIALLLLL}
  \Lambda^{MN} \ = \ \partial^{[M}{}_{K}\chi^{N]K}\;, \qquad 
  \Sigma_{MN} \ = \ -\frac{1}{4} \partial_{MN}\partial_{KL}\chi^{KL}\;. 
 \ee 
Again,  triviality follows from the section constraints, which immediately imply that transport (and density) terms 
vanish, as well as $K^{MN}=0$ by a quick computation with (\ref{simplifiedSec}). 
Note that $\chi^{MN}$ can be symmetric, in which case the $\Sigma$ parameter vanishes. 
In particular, this contains as a special case the familiar 
DFT trivial parameter $\Lambda^{MN}=\partial^{MN}\chi$ via $\chi^{MN}\equiv \chi\,\eta^{MN}$. 
There is a more general trivial parameter for the latter: 
 \be
  \Lambda^{MN} \ = \ \Omega^{MN}\;, \quad \text{with $\Omega^{MN}$ covariantly constrained}\,,  
 \ee
by which we mean $\Omega^{MN}\partial_{MN}=0$, etc.  
Finally, there is a trivial parameter that generalizes (\ref{TRIVIALLLLL}) for $\chi^{MN}$ antisymmetric.  
Indeed, the E$_{8(8)}$ case suggests that 
$\Lambda^{MN}=f^{MN,KL}{}_{PQ}\Omega_{KL}{}^{PQ}$, where $\Omega$ is covariantly constrained 
in the first index, is trivial. Here we find that 
 \be\label{NEWtriVIAL}
 \begin{split}
  &\Lambda^{MN} \ = \ \Omega^{[M}{}_{K}{}^{N]K}\;, \qquad \Sigma_{MN} \ = \
   -\frac{1}{8}\partial_{MN}\Omega_{KL}{}^{KL} - \frac{1}{8}\partial_{KL}\Omega_{MN}{}^{KL}\\
   &\text{with $\,\Omega_{KL}{}^{PQ}\,$ \textit{covariantly constrained} (\ref{sectionSigma}) in the first index pair} \,, 
 \end{split}
 \ee  
 is indeed trivial.

\subsection{Generalized Dorfman structure}\label{genDorfmanSEC} 

 We will now introduce a new notation that allows us to exhibit an algebraic structure on the 
 space of gauge parameters $\Lambda^{MN}$, $\Sigma_{MN}$ that is analogous to the 
 Dorfman product appearing for DFTs and ExFTs with external dimension 
$n\geq 4$. 
We introduce for the gauge parameters the pair notation 
 \be\label{Upsilon}
  \Upsilon \ \equiv  \ (\Lambda,\Sigma)\;, 
 \ee
and we treat the first entry as an adjoint vector $\Lambda^{MN}$ of weight  $\lambda=1$ 
and the second entry as a co-adjoint vector $\Sigma_{MN}$ of weight zero that is covariantly 
constrained according to (\ref{sectionSigma}).\footnote{Of course, since we have a metric to raise and lower 
indices,  adjoint and co-adjoint representations are actually equivalent, but it is sometimes useful 
to make this distinction in order to keep track of the two components.}

Our goal is to define a product for such doubled objects such that its antisymmetric part coincides 
with the gauge algebra structure introduced in the previous subsection and its symmetric part is a trivial gauge parameter. 
It turns out these conditions are satisfied for 
 \be\label{DorfmanNN}
 \begin{split}
   \Upsilon_1 \circ \Upsilon_2 \ &\equiv \ (\Lambda_1,\Sigma_1)\circ (\Lambda_2,\Sigma_2) \\
    \ &\equiv \ 
  \Big(\,{\cal L}^{[1]}_{\Upsilon_1}\Lambda_2^{MN}\,, \;
    {\cal L}^{[0]}_{\Upsilon_1}\Sigma_{2MN}
    +\frac{1}{4} \Lambda_2^{KL}\partial_{MN}K(\Upsilon_1)_{KL}\, \Big)\;, 
 \end{split}   
 \ee  
where we used the notation (\ref{KDEFF}) for $K(\Upsilon_1)\equiv K(\Lambda_1,\Sigma_1)$.  
Moreover, the Lie derivatives in here act as defined in the previous subsection, with 
$\Lambda$ carrying weight one and $\Sigma$ weight zero.  
Specifically, using that $\Sigma$ is constrained one computes  
 \be
   {\cal L}_{\Upsilon_1}\Sigma_{2MN} \ = \ {\cal L}_{(\Lambda_1,\Sigma_1)}^{[0]}\Sigma_{2MN} \ = \ 
  \Lambda_1^{KL}\partial_{KL}\Sigma_{2MN} + \partial_{MN}\Lambda_1^{KL} \Sigma_{2KL}
  +\partial_{KL}\Lambda_1^{KL} \Sigma_{2MN}\;. 
 \ee  
Note that, curiously,  the `anomalous' terms in the $\Sigma$ component of (\ref{DorfmanNN})  
have the order of 1 and 2 such that 
we \textit{cannot} think of this as a deformed Lie derivative of $\Sigma_2$ w.r.t.~$\Upsilon_1$, 
because $\Lambda_2$ enters explicitly. This ordering  
turns out to be crucial for the following construction.  

We first verify that the antisymmetrized product defines the expected bracket:  
 \be\label{bracketISantisymm}
  \big[\Upsilon_1,\Upsilon_2\big] \ \equiv \ \frac{1}{2}( \Upsilon_1 \circ \Upsilon_2 - \Upsilon_2 \circ \Upsilon_1)
   \ \equiv \ \big[(\Lambda_1,\Sigma_1),(\Lambda_2,\Sigma_2)\big]
  \ \equiv \  (\Lambda_{[1,2]}, \Sigma_{[1,2]})
  \;, 
 \ee 
where 
 \be
 \begin{split}
   \Lambda_{[1,2]}^{MN} \ \equiv \ &\, \Lambda_{1}{}^{KL}\partial_{KL}\Lambda_{2}{}^{MN}
   -3\,\Lambda_{1}{}^{[MN}\partial_{KL}\Lambda_{2}{}^{KL]}
   -2\,\Lambda_{1K}{}^{[M}\partial_L{}^{N]}\Lambda_{2}{}^{KL} \\
   &\,+4\,\Sigma_{1}^{[M}{}_K \Lambda_2^{|K|N]} - (1\leftrightarrow 2)\;, \\
   \Sigma_{[1,2]MN} \ \equiv \ &\,  
   \frac{1}{2}(\Lambda_1^{KL}\partial_{KL}\Sigma_{2MN} + \partial_{MN}\Lambda_1^{KL} \Sigma_{2KL}
   +\partial_{KL}\Lambda_1^{KL} \Sigma_{2MN}\\
   &\;\; \; - \Lambda_1^{KL}\partial_{MN} \Sigma_{2KL}
  - \Lambda_{1}{}^{P}{}_{K}\partial_{MN}\partial_{PL}\Lambda_{2}{}^{KL}
  - (1\leftrightarrow 2))\;. 
 \end{split}
 \ee  
This is not quite of the form (\ref{simplifiedGaugeAlgebra}), but is equivalent to it upon adding 
trivial gauge parameters. Indeed, the gauge algebra is only well-defined up to trivial gauge parameters, 
and adding a trivial parameter  of the form (\ref{NEWtriVIAL}), with 
 \be
  \Omega_{MN}{}^{KL} \ = \ -4\,\Sigma_{1MN}\Lambda_2^{KL}  - (1\leftrightarrow 2)\;, 
 \ee
shows that the above indeed defines the gauge algebra bracket. 
Next we have to prove that the symmetric  part of the product is trivial. 
We compute: 
  \be\label{bracketISsymm}
  \begin{split}
  \frac{1}{2}( \Upsilon_1 \circ \Upsilon_2 + \Upsilon_2 \circ \Upsilon_1) \ = \ 
  \Big(\, &3\,\partial_{KL}\big(\Lambda_2{}^{[MN}
  \Lambda_1{}^{KL]}\big)+\Omega^{[M}{}_{K}{}^{N]K}+\partial^{[M}{}_{K}\chi^{N]K}\, , \; \\
   &\qquad -\frac{1}{8}\partial_{MN}\Omega_{KL}{}^{KL} - \frac{1}{8}\partial_{KL}\Omega_{MN}{}^{KL} 
 \Big)\;, 
 \end{split}
 \ee 
where 
 \be
 \begin{split}
  \Omega_{MN}{}^{KL} \ &\equiv \ -4\,\Sigma_{1MN}\Lambda_2{}^{KL} - 2\,\partial_{MN}\Lambda_{1}{}^{[K}{}_{P}
  \Lambda_2{}^{L]P}+(1\leftrightarrow 2)\;, \\
  \chi^{MN} \ &\equiv \ 2\,\Lambda_2{}^{(M}{}_{K}\, \Lambda_1{}^{N)K}\;. 
 \end{split}
 \ee 
We infer that the result is indeed of the trivial form (\ref{Trivial1}), (\ref{TRIVIALLLLL}) and (\ref{NEWtriVIAL}).

Our next goal is to show that the product satisfies a certain Jacobi or Leibniz-type identity that will 
be instrumental for our subsequent construction. To this end it is convenient to extend the notion 
of generalized Lie derivative slightly so as to act on doubled objects 
${\mathfrak A}\equiv ({\cal A}^{MN}, {\cal B}_{MN})$ 
of the same type as $\Upsilon$: 
 \be\label{genegenLIE}
  {\cal L}_{\Upsilon}{\mathfrak A} \ \equiv \ \Upsilon\circ {\mathfrak A}\;. 
 \ee 
From the definition (\ref{DorfmanNN}) of the product we infer that for the first component (the 
`$\Lambda$ component')  
this reduces to the conventional generalized Lie derivative, but for the $\Sigma$ component there 
is an additional contribution due to the `anomalous' term in (\ref{DorfmanNN}). 
We will next prove, however, that these extended generalized Lie derivatives still close according to 
the same bracket: 
\be\label{BracketUpsilons}
  \big[{\cal L}_{\Upsilon_1}, {\cal L}_{\Upsilon_2}\big]\,{\mathfrak A} \ = \  
  {\cal L}_{[\Upsilon_1,\Upsilon_2]}{\mathfrak A} \;. 
 \ee 
Again, for the $\Lambda$ component this reduces to the closure of standard generalized 
Lie derivatives established in the previous subsection, 
but for the $\Sigma$ component one obtains additional contributions, so that after a brief computation 
 \be
  \big[{\cal L}_{\Upsilon_1}, {\cal L}_{\Upsilon_2}\big]{\mathfrak A} \ = \ 
  \Big({\cal L}_{[\Upsilon_1,\Upsilon_2]}{\cal A}^{MN}\;, \; {\cal L}_{[\Upsilon_1,\Upsilon_2]}{\cal B}_{MN}
  +\frac{1}{4}{\cal A}^{KL}{\cal L}_{\Upsilon_1}(\partial_{MN}K(\Upsilon_2)_{KL}) -(1\leftrightarrow 2)\, \Big)\;. 
 \ee 
On the other hand, the right-hand side of (\ref{BracketUpsilons}) equals 
\be
  {\cal L}_{[\Upsilon_1,\Upsilon_2]}{\mathfrak A} \ = \ 
  \Big({\cal L}_{[\Upsilon_1,\Upsilon_2]}{\cal A}^{MN}\;, \; {\cal L}_{[\Upsilon_1,\Upsilon_2]}{\cal B}_{MN}
  +\frac{1}{4}{\cal A}^{KL}\partial_{MN}K([\Upsilon_1,\Upsilon_2])_{KL}\, \Big)\;. 
 \ee 
In order to prove that the above two right-hand sides are actually identical  we use 
 \be
  \partial_{MN}({\cal L}_{\Upsilon_1}K(\Upsilon_2)_{KL}) \ = \ 
  {\cal L}_{\Upsilon_1}(\partial_{MN}K(\Upsilon_2)_{KL})
  +2\,\partial_{MN}K(\Upsilon_1)_{[K}{}^{P} K(\Upsilon_2)_{|P|L]}\;. 
 \ee 
This follows as in (\ref{noncovVAR}), using that the Lie derivative acts on $K$, defined in (\ref{KDEFF}), 
as a tensor of zero density weight.  
With this one can quickly establish 
 \be
 \begin{split}
  \big[{\cal L}_{\Upsilon_1}, {\cal L}_{\Upsilon_2}\big]\,{\mathfrak A} \ = \  
  {\cal L}_{[\Upsilon_1,\Upsilon_2]}{\mathfrak A} 
  +\Big(0\;,\; \frac{1}{4} {\cal A}^{KL}\partial_{MN}X_{KL}
  \, \Big)\;, 
 \end{split}
 \ee 
where 
 \be
  X_{KL} \ \equiv \  {\cal L}_{\Upsilon_1}K(\Upsilon_2)_{KL}
  -K(\Upsilon_1)_K{}^{P} K(\Upsilon_2)_{PL}-(1\leftrightarrow 2) -K([\Upsilon_1,\Upsilon_2])_{KL}\;. 
 \ee
Using (\ref{KClosureREL}) it is easy to see that this is actually zero, completing the proof of (\ref{BracketUpsilons}).

We now derive a Leibniz identity for the product from the closure relation (\ref{BracketUpsilons}). 
We first note that for $\Upsilon$ trivial the extended generalized Lie derivative (\ref{genegenLIE}) vanishes: 
 \be\label{trivialtrivialproduct}
  \Upsilon \quad {\rm trivial}\qquad \Rightarrow \qquad \Upsilon \circ {\mathfrak A} \ = \ 0\;. 
 \ee  
 This holds by definition for the $\Lambda$ component and for the $\Sigma$ component 
follows from the fact that the $K(\Upsilon)$ entering the anomalous term of (\ref{DorfmanNN}) 
is zero for trivial parameters. 
Thus, using that the symmetric part (\ref{bracketISsymm}) of the product is trivial, 
the closure relation can also be written as 
  \be\label{commisDORFMAN}
  \big[{\cal L}_{\Upsilon_1}, {\cal L}_{\Upsilon_2}\big]\,{\mathfrak A} \ = \  
  {\cal L}_{\Upsilon_1\circ \Upsilon_2}{\mathfrak A} \;. 
 \ee 
Using (\ref{genegenLIE}) twice we can write this as 
 \be
  \Upsilon_1\circ (\Upsilon_2\circ {\mathfrak A}) -   \Upsilon_2\circ (\Upsilon_1\circ {\mathfrak A})
  \ = \ (\Upsilon_1\circ \Upsilon_2)\circ {\mathfrak A}\;. 
 \ee 
Upon renaming the doubled objects entering here and reordering the equations, we have thus established  
the Leibniz identity
  \be\label{LEIBNIZ}
   {\mathfrak A}\circ ({\mathfrak B}\circ {\mathfrak C}) 
   \ = \ ({\mathfrak A}\circ {\mathfrak B})\circ {\mathfrak C} +{\mathfrak B}\circ ({\mathfrak A}\circ {\mathfrak C})\;. 
  \ee 
Let us finally note that formally all relations that hold for conventional Dorfman products are then 
also satisfied for the product defined here, except that the relevant objects are doubled in the 
sense of (\ref{Upsilon}). In particular, the Jacobiator of the bracket (\ref{bracketISantisymm}) can 
then be proved to be trivial in precise analogy to the original DFT and ExFTs for E$_{n(n)}$ with 
$n\leq 7$.

\subsection{Gauge fields, tensor hierarchy, and Chern-Simons action} 

We will now introduce gauge fields that, roughly speaking, take values in the 
algebra given by the Dorfman product defined above.  
More precisely, we introduce gauge fields ${\cal A}_{\mu}{}^{MN}$ and ${\cal B}_{\mu\, MN}$ and combine 
them into a pair or doubled object as above: 
 \be\label{DoubledVEctor}
  \mathfrak{A}_{\mu} \ \equiv \ \big({\cal A}_{\mu}{}^{MN}, {\cal B}_{\mu\, MN}\big)\;.  
 \ee 
In particular, ${\cal A}$ carries weight one and ${\cal B}$ weight zero while being constrained
according to (\ref{sectionSigma}), i.e.
 \bea
{\cal B}_{\mu\, [MN}\otimes \partial_{KL]} \ = \  0  \ = \  \eta^{NK}\,{\cal B}_{\mu\, MN}\otimes \partial_{KL}
\;,\qquad \mbox{etc.}\;.
\label{sectionB}
\eea
Their transformation rules receive 
inhomogeneous terms as to be expected for gauge fields. 
Indeed, in analogy to Yang-Mills theories we postulate the following gauge transformations 
w.r.t.~doubled parameters (\ref{Upsilon}) 
 \be\label{connTRANs}
  \delta_{\Upsilon} \mathfrak{A}_{\mu}   \ \equiv \ \mathfrak{D}_{\mu}\Upsilon  \;, 
 \ee 
where we defined the covariant derivative 
 \be\label{covDEFfrak}
  \mathfrak{D}_{\mu} \ = \ \partial_{\mu} \ - \ \mathfrak{A}_{\mu}\, \circ \;. 
 \ee 
It should be emphasized that the covariant derivative as written is only defined 
on doubled objects, which is indicated by the mathfrak notation. 
We can, however, define covariant derivatives for any field with a 
well-defined action of the generalized Lie derivatives in sec.~\ref{firstsubsection}. 
For a generic (undoubled) tensor field $V$ we define 
 \be\label{genericcovDER}
  D_{\mu}V \ \equiv \ \partial_{\mu}V - {\cal L}_{({\cal A}_{\mu}, {\cal B}_{\mu})}V\;. 
 \ee 
For instance, for a vector $V^M$ of zero weight this reads explicitly 
\bea
D_\mu V^{M} &\equiv&
\partial_\mu V^{M}
-{\cal A}_\mu{}^{KL}\partial_{KL}\,V^{M}
+2\, \left(\partial^{MP}{\cal A}_{\mu\,PK}-\partial_{KP}{\cal A}_\mu{}^{PM}\right) V^{K}
\nonumber\\
&&{}
-4\,\eta^{ML} \,{\cal B}_{\mu\,LK}\,  V^{K}
\;.
\eea 
Despite $V$ not being a doubled object we can prove in an index-free fashion 
that the covariant derivative indeed transforms covariantly: 
 \be 
  \begin{split}
   \delta_{(\Lambda,\Sigma)}(D_{\mu} V) \ &= \ \delta_{\Upsilon}(\partial_{\mu}V-{\cal L}_{\mathfrak{A}_{\mu}}V)
   \ = \ \partial_{\mu}({\cal L}_{\Upsilon}V) - {\cal L}_{\partial_{\mu}\Upsilon-\mathfrak{A}_{\mu}\circ \Upsilon}V
   -{\cal L}_{\mathfrak{A}_{\mu}}({\cal L}_{\Upsilon}V) \\
   \ &= \ {\cal L}_{\partial_{\mu}\Upsilon}V+{\cal L}_{\Upsilon}(\partial_{\mu}V)
   -{\cal L}_{\partial_{\mu}\Upsilon}V + {\cal L}_{\mathfrak{A}_{\mu}\circ \Upsilon}V
   -{\cal L}_{\mathfrak{A}_{\mu}}({\cal L}_{\Upsilon}V) \\
   \ &= \ {\cal L}_{\Upsilon}(\partial_{\mu}V-{\cal L}_{\mathfrak{A}_{\mu}}V)
   +([{\cal L}_{\Upsilon},{\cal L}_{\mathfrak{A}_{\mu}}] + {\cal L}_{\mathfrak{A}_{\mu}\circ \Upsilon})V \\
   \ &= \ {\cal L}_{\Upsilon}(D_{\mu}V)\;, 
  \end{split}
 \ee  
where we used (\ref{commisDORFMAN}) in the last step. 
This proves the covariance of the covariant derivative under combined tensor transformations given 
by generalized Lie derivatives and vector gauge transformations, whose component form is  
with (\ref{connTRANs}) and  (\ref{DorfmanNN}) found to be 
 \be
  \begin{split}
   \delta_{(\Lambda,\Sigma)}{\cal A}_{\mu}{}^{MN} \ &= \ D_{\mu}\Lambda^{MN}\;, \\
   \delta_{(\Lambda,\Sigma)}{\cal B}_{\mu \, MN} \ &= \ D_{\mu}\Sigma_{MN}
   -\Lambda^{KL}\partial_{MN} {\cal B}_{\mu KL} - \Lambda^{K}{}_{L}\partial_{MN}\partial_{KP}
   {\cal A}_{\mu}{}^{LP}\;, 
  \end{split}
 \ee
which of course may also be verified with a direct component computation.    
This clarifies the seemingly `non-covariant' terms in the gauge transformations of ${\cal B}_{\mu}$, 
first identified for the SL$(2,\mathbb{R})$ and E$_{8(8)}$ cases \cite{Hohm:2013jma,Hohm:2014fxa}, 
and explains them as a consequence 
of the  necessary `anomalous' terms of the Dorfman product.

Let us next discuss the gauge structure and invariant field strengths for the gauge vectors. 
With the Leibniz identity (\ref{LEIBNIZ}) it is straightforward to compute the commutator of two 
gauge transformations (\ref{connTRANs}): 
  \be\label{almostCLOSURE}
   \big[\delta_{\Upsilon_1},\delta_{\Upsilon_2}\big]\, \mathfrak{A}_{\mu} \ = \ 
   \delta_{\Upsilon_1\circ \Upsilon_2}\mathfrak{A}_{\mu} + 2\, \{ \Upsilon_{[1},\mathfrak{D}_{\mu}\Upsilon_{2]}
    \} \;, 
  \ee 
where we introduced the notation  
  \be
   \{{\mathfrak A}, {\mathfrak B}\} \ \equiv \ {\mathfrak A}\circ {\mathfrak B} \ + \ {\mathfrak B}\circ {\mathfrak A}\;. 
  \ee 
We infer from (\ref{almostCLOSURE}) that the vector gauge transformations do not quite close, 
but the failure of closure involves the symmetrized product, which is trivial, c.f.~(\ref{bracketISsymm}). 
This implies that the extra terms can be absorbed into higher-form (here 2-form) gauge transformations, 
as is standard in the tensor hierarchy. Thus, the combined one- and two-form transformations close. 
Another way to see the need for 2-forms is by inspection of the naive field strength for the gauge vectors: 
  \be\label{frakFfieldstrength}
   \mathfrak{F}_{\mu\nu} \ \equiv \ \partial_{\mu}\mathfrak{A}_{\nu}
   \ - \  \partial_{\nu}\mathfrak{A}_{\mu} \ - \ \big[ \mathfrak{A}_{\mu},\mathfrak{A}_{\nu}\big] \ + \ \cdots \;, 
  \ee 
 where the ellipsis  denotes 2-form terms to be added  momentarily. 
 In components, writing $\mathfrak{F}_{\mu\nu}\equiv (F_{\mu\nu}, G_{\mu\nu})+\cdots$, this reads 
 \bea
F_{\mu\nu}{}^{MN} &\equiv&
2\,\partial_{[\mu}{\cal A}_{\nu]}{}^{MN}-
2\,{\cal A}_{[\mu}{}^{KL}\,\partial_{KL}{ \cal A}_{\nu]}{}^{MN}
+6\,{\cal A}_{[\mu}{}^{[MN} \partial_{KL}{\cal A}_{\nu]}{}^{KL]}
-4\,{\cal A}_{[\mu}{}^{K[M}
\partial^{N]L}{\cal A}_{\nu]\,KL}
,\nonumber\\
G_{\mu\nu\,MN} &\equiv&
2\,D_{[\mu} {\cal B}_{\nu]}{}_{MN}
-{\cal A}_{[\mu\,K}{}^{P} \,
 \partial_{PQ} \partial_{MN}{\cal A}_{\nu]}{}^{KQ}
 \;.
\eea
 We consider now the general variation under an arbitrary $\delta \, \mathfrak{A}_{\mu}$, 
 for which we compute 
 \be 
  \delta\, \mathfrak{F}_{\mu\nu} \ = \ 2\,\mathfrak{D}_{[\mu}\, \delta \, \mathfrak{A}_{\nu]} + \{ \mathfrak{A}_{[\mu}, 
  \delta \, \mathfrak{A}_{\nu]}\} + \cdots \;. 
 \ee   
We do not quite obtain the expected identity with only the covariant curl of $\delta \, \mathfrak{A}_{\mu}$, 
but the additional terms are trivial and can hence 
be absorbed into the 2-forms. More precisely, 2-forms are introduced in precise correspondence 
with the trivial terms in the symmetrized product (\ref{bracketISsymm}). We thus define the full field strength to be 
$\mathfrak{F}_{\mu\nu }\equiv ({\cal F}_{\mu\nu}, {\cal G}_{\mu\nu})$, where 
\bea
{\cal F}_{\mu\nu}{}^{MN} &=& F_{\mu\nu}{}^{MN} 
+ \partial_{KL}{C}_{\mu\nu}{}^{[KLMN]}
+\partial_{K}{}^{[M}\,{C}_{\mu\nu}{}^{N]K}
+8\, {\cal C}_{\mu\nu\,KL}{}^{K[M}\,\eta^{N]L} 
\;,
\nonumber\\
{\cal G}_{\mu\nu\,MN} &=& G_{\mu\nu\,MN} +
\partial_{KL}{\cal C}_{\mu\nu\,MN}{}^{KL} +
\partial_{MN}{\cal C}_{\mu\nu\,KL}{}^{KL}
\;,
\label{fullFG}
\eea
and the two-forms ${\cal C}_{\mu\nu\,MN}{}^{KL}$
are covariantly constrained in its indices $[MN]$\,.
After adding the appropriate 2-forms to the field strength, we can 
show its complete gauge covariance. To this end, we use the identity  
 \be\label{extendedcomm}
  \big[\mathfrak{D}_{\mu}, \mathfrak{D}_{\nu}\big]\Upsilon \ = \ -\mathfrak{F}_{\mu\nu}\circ \Upsilon \;, 
 \ee  
which follows immediately from (\ref{covDEFfrak}) and the fact that the 2-form contributions are of the trivial form
and hence 
drop out of this relation by (\ref{trivialtrivialproduct}).  We similarly have for the covariant derivatives (\ref{genericcovDER}) 
\bea{}
\big[D_\mu,D_\nu\big] V^{M} &=&
-{\cal L}_{(F_{\mu\nu},G_{\mu\nu})}\,V^M
~=-{\cal L}_{({\cal F}_{\mu\nu},{\cal G}_{\mu\nu})}\,V^M
\;. 
\eea
This is contained in (\ref{extendedcomm}), which can be evaluated on the first $(\Lambda)$ component of a 
doubled object, thereby reproducing this equation.  
We then compute with (\ref{extendedcomm}) 
  \be\label{fieldstrengthsgaugevar}
   \delta_{\Upsilon}\mathfrak{F}_{\mu\nu} 
      \ = \ \Upsilon\circ \mathfrak{F}_{\mu\nu}\;, 
  \ee 
using that up to trivial contributions taken care of by the 2-forms the order of the product can be exchanged 
up to a sign.

\subsubsection*{Chern-Simons term}
We will now define a Chern-Simons action for the gauge vectors $\mathfrak{A}_{\mu}$. 
To this end we need an invariant inner product. 
 The naive ansatz for the `off-diagonal' inner product between adjoint and co-adjoint vector needs to be deformed by a derivative term in order to 
 account for the `anomalous' term in the $\Sigma$ component of the product. One finds: 
  \be
   \langle \langle {\mathfrak A}_1, {\mathfrak A}_{2}\rangle\rangle  \ \equiv \ \langle \langle ({\cal A}_1,{\cal B}_1), 
   ({\cal A}_2,{\cal B}_2)
   \rangle\rangle  \ \equiv \ 
   2\, {\cal A}_{(1}{}^{MN} {\cal B}_{2)MN} + {\cal A}_{(1}{}^{MN}\partial_{MK} {\cal A}_{2)N}{}^{K}\;. 
   \label{prod}
  \ee  
 The invariance condition means, more precisely, invariance up to total derivatives: 
  \be\label{COVcondINNER}
   \langle \langle \Upsilon\circ {\mathfrak A}_1,{\mathfrak A}_2\rangle \rangle 
   +\langle\langle  {\mathfrak A}_1,\Upsilon\circ {\mathfrak A}_2\rangle \rangle 
   \ = \ \partial_{MN}(\Lambda^{MN} \langle\langle  {\mathfrak A}_1, {\mathfrak A}_{2}\rangle \rangle)\;, 
  \ee 
which can be verified by an explicit computation.  
Thus, a truly invariant inner product involves the $N$-dimensional $Y$ integration (where
$N=\frac{1}{2}(p+q)(p+q-1))$: 
 \be\label{FinalINV}
  \big\langle   {\mathfrak A}_1, {\mathfrak A}_{2}\big\rangle \ \equiv \ 
   \int {\rm d}^NY \big(  {\cal A}_1{}^{MN} {\cal B}_{2MN} +{\cal A}_2{}^{MN} {\cal B}_{1MN} 
   +{\cal A}_{1}{}^{MN}\partial_{MK}{\cal A}_{2N}{}^{K}\big)\;, 
  \ee 
where we used that one can integrate by parts in the terms with derivatives 
to combine two terms into one. We can then also write, using the notation (\ref{KDEFF}),  
 \be\label{innerintegrated}
    \big\langle   {\mathfrak A}_1, {\mathfrak A}_{2}\big\rangle \ = \ 
   \int {\rm d}^NY \big(\,\frac{1}{4} {\cal A}_1^{MN} K_{MN}({\mathfrak A}_2) 
   + {\cal A}_2^{MN} {\cal B}_{1MN}\, \big)\;. 
 \ee  
Although no longer manifest, the inner product defined in this way is of course still symmetric in the two arguments, up to boundary terms. 
An important consequence is that  the inner product is zero 
whenever one argument is trivial: 
 \be\label{degeneracy}
  {\mathfrak T}\quad {\rm trivial} \qquad \Rightarrow \qquad \big\langle {\mathfrak A}\,,
  {\mathfrak T}\big\rangle   \ = \ 0
  \quad \forall \; {\mathfrak A}\;. 
 \ee 
This follows directly from (\ref{innerintegrated}), 
 \be\label{triviadirection}
  \big\langle {\mathfrak A}\, ,{\mathfrak T}\big\rangle  \ = \ 
  \int {\rm d}^NY \big(\,\frac{1}{4} {\cal A}^{MN} K_{MN}({\mathfrak T}) 
  + {\cal T}^{MN} {\cal B}_{MN}\, \big) \ = \ 0\;, 
 \ee
using that for  trivial ${\mathfrak T}=({\cal T}^{MN},\tau_{MN})$ we have $K({\mathfrak T})=0$ 
and that the contraction of ${\cal T}^{MN}$ with any covariantly constrained object such as 
${\cal B}_{MN}$ vanishes.

 Having established the existence of an invariant inner product, a natural ansatz 
for the Chern-Simons action is its familiar three-dimensional form: 
  \be
  S_{\rm CS} \ = \ \int {\rm d}^3x \,\varepsilon^{\mu\nu\rho}\big(  \big\langle\, \mathfrak{A}_{\mu}\,,\,
  \partial_{\nu}\mathfrak{A}_{\rho} \,\big\rangle  -\frac{1}{3} \big\langle\, \mathfrak{A}_{\mu}\,,\, 
   \mathfrak{A}_{\nu}\circ \mathfrak{A}_{\rho}\,\big\rangle \big)\;, 
   \label{SCS}
 \ee 
where the internal integration is implicit in the inner product.  
Using the Leibniz identity, its gauge variation up to total derivatives can be written as 
 \be
  \delta_{\Upsilon}S_{\rm CS} \ = \ -\frac{2}{3}\,\int {\rm d}^3x \, 
  \varepsilon^{\mu\nu\rho}\, \big\langle \mathfrak{A}_{\mu}, 
  \{\mathfrak{A}_{\nu},\partial_{\rho}\Upsilon\}\big\rangle \ = \ 0 \;,  
 \ee
which vanishes as a consequence of (\ref{degeneracy}) since the symmetric pairing $\{\,,\,\}$ is trivial. 
 Using the Leibniz identity (\ref{LEIBNIZ}) again, 
 one can show that under an arbitrary variation $\delta\,\mathfrak{A}_{\mu}$ 
  \be\label{generalCSvar}
   \delta S_{\rm CS} \ = \ \int {\rm d}^3x\,\varepsilon^{\mu\nu\rho}\big\langle \delta\,\mathfrak{A}_{\mu}\, , \,  
   \mathfrak{F}_{\nu\rho}\big\rangle \;. 
  \ee
 Because of the degeneracy (\ref{degeneracy}) this does not imply that the field equations are 
 $\mathfrak{F}_{\mu\nu}=0$, but only a suitably projected version of the field strength is zero. 
 In the following we will couple such a Chern-Simons action to charged matter, such that the field
 equations relate a projection of the field strength to scalar currents.
 We can now use this result 
 to compare with the more familiar form of this variation. We first recall the identification 
  \be 
   \mathfrak{F}_{\mu\nu} \ = \ \big({\cal F}_{\mu\nu}{}^{MN}, {\cal G}_{\mu\nu MN}\big)\;. 
  \ee
We then read off from (\ref{generalCSvar}) and (\ref{FinalINV}) 
 \be\label{covCSvar}
  \delta S_{\rm CS} \ = \ \int {\rm d}^3x\,{\rm d}^NY\,\varepsilon^{\mu\nu\rho}\Big( 
  \delta {\cal B}_{\mu MN} {\cal F}_{\nu\rho}{}^{MN} + \delta {\cal A}_{\mu}{}^{MN}
  \big({\cal G}_{\nu\rho \, MN}+\partial_{MK} {\cal F}_{\nu\rho N}{}^{K}\big)\Big)\;. 
 \ee

\subsection{Covariant derivatives and variations} 

For completeness and in order to relate to the `covariant variations' employed for the supersymmetric 
E$_{8(8)}$ ExFT in \cite{Baguet:2016jph}, we will now discuss some aspects of the `${\rm O}(p,q)$ 
covariant' geometry, notably the notion of connections and torsion. 
We begin by introducing derivatives that covariantize the internal partial derivatives w.r.t.~the 
internal generalized diffeomorphisms. For a (co-adjoint) vector of weight zero we define  
  \be
   \nabla_{\cal M}V_{\cal N} \ \equiv \ \partial_{\cal M} V_{\cal N} 
   - \Gamma_{{\cal M},{\cal N}}{}^{\,\cal K} \, V_{\cal K}\;, 
  \ee
with connections $\Gamma_{{\cal M},{\cal N}}{}^{\,\cal K}$ that take values in the Lie algebra 
$\mathfrak{so}(p,q)$. We can thus introduce $\Gamma_{{\cal M}, {\cal N}}$ by 
 \be
  \Gamma_{{\cal M},{\cal N}}{}^{\, \cal K} \ \equiv \
  \Gamma_{{\cal M},{\cal L}} \, f^{{\cal L}{\cal K}}{}_{\cal N}\;, 
   \ee  
which reads  in index pairs
  \be
   \Gamma_{MN,KL}{}^{PQ} \ = \ \frac{1}{4}\, \Gamma_{MN,RS}\,f^{RS,PQ}{}_{KL}\;, 
  \ee
 with a pre-factor for later convenience.  This implies for fundamental vectors  
 \be
 \begin{split}
  \nabla_{MN} V_{K} \ &= \ \partial_{MN} V_{K}-\Gamma_{MN,K}{}^{L} V_{L}\;, \\
  \nabla_{MN} V^{K} \ &= \ \partial_{MN} V^{K}+\Gamma_{MN,L}{}^{K} V^{L}\;. 
 \end{split} 
 \ee     
In (\ref{noncovVAR}) we computed the non-covariant gauge transformation for a 
partial derivative of a vector. From this result and the first equation above 
we infer that the covariant derivative indeed 
transforms covariantly  provided 
the connection transforms as tensor of weight $\lambda=-1$, plus the usual 
inhomogeneous term: 
 \be\label{conntransform}
  \delta_{\Upsilon}\Gamma_{MN,KL}  \ = \ 
  \partial_{MN}K_{KL}(\Upsilon) +{\cal L}^{[-1]}_{\Upsilon}\Gamma_{MN,KL} \;,
 \ee
 with gauge parameter (\ref{Upsilon}), and $K_{KL}$ defined in (\ref{KDEFF}).
We can also define the covariant derivative of a tensor of arbitrary density weight $\lambda$, 
using that the above implies for the non-covariant variation 
 \be
  \Delta_{\Lambda}^{\rm nc}\big(\Gamma_{[M}{}^{K}{}_{,N]K}\big) \ = \ 
  \partial_{MN}(\partial_{KL}\Lambda^{KL})\;. 
 \ee   
Thus, for a vector of weight $\lambda$, 
 \be
  \nabla_{MN} V_{K} \ = \ \partial_{MN} V_{K}-\Gamma_{MN,K}{}^{L} V_{L}
  -\lambda\,\Gamma_{[M}{}^{L}{}_{,N]L}\, V_K\;. 
 \ee
  
\medskip

We next aim to define a torsion tensor as a particular projection of the 
connection that transforms tensorially. 
In general, the connection lives in the tensor product 
\Yboxdim{0.9em}
\bea
 \Gamma_{MN,KL}\;:\qquad 
 \Yvcentermath1 \gyoung(;,;) \;\otimes\; \gyoung(;,;) &=&
 \Yvcentermath1  
 \gyoung(;,;,;,;)  \;\oplus \; \gyoung(;;,;,;) \;\oplus \; \gyoung(;,;) \;\oplus \; \gyoung(;;,;;)  \;\oplus \; \gyoung(;;)  \;\oplus \; \bullet
 \;\;, 
 \label{tensprod}
\eea 
where the Hook and window tableaux are traceless, with the antisymmetric and symmetric  tableaux 
carrying two boxes denoting their trace parts. The latter is traceless itself with its trace give by the singlet $\bullet$\,.
We next use that the section constraint implies 
 \be 
  \partial_{[MN}K_{KL]} \ =\  0\;, \qquad \partial_{(M}{}^{K}K_{N)K}\ = \ 0\;, 
\ee 
as may be quickly verified by a direct computation. 
We then infer with (\ref{conntransform}) that the following projections have tensor character: 
 \be
  {\cal T}_{MNKL} \ \equiv \ 6\, \Gamma_{[MN,KL]}\;, \qquad
  {\cal T}_{MN} \ \equiv \ 2\,\Gamma_{(M}{}^{K}{}_{,N)K} \;,  
 \ee 
corresponding to the totally antisymmetric and the symmetric trace tableaux. We may also combine this into 
a reducible torsion tensor: 
 \be\label{FULLTorsion}
  {\cal T}_{MN,KL} \ \equiv \ {\cal T}_{MNKL} + 2\,{\cal T}_{[M [\underline{K}}\, \eta_{N]\underline{L}]}\;. 
 \ee 
In the following, we will thus impose torsionlessness of the connection $\Gamma$ as
\bea
 {\cal T}_{MN,KL} &=& 0 \;.
 \label{T0}
\eea
As usual in generalized geometries, this condition does not fully determine the connection 
\cite{Siegel:1993th}
but all the parts that are required in order to formulate the field equations and transformation rules.
For DFTs and ExFTs with external dimension $n\geq 4$ the torsion tensor is such that for 
vanishing torsion the manifestly 
covariant Lie derivative in which all partial derivatives have been replaced by covariant derivatives 
equals the original generalized Lie derivative. The same is not quite true for 
ExFTs with $n=3$~\cite{Cederwall:2015ica,Baguet:2016jph}, but we have the following close analogue: For 
 \be
  \widetilde{\Sigma}_{MN} \ \equiv \ \Sigma_{MN} -\frac{1}{4}\,\Gamma_{MN,KL}\, \Lambda^{KL}\;, 
 \ee
one can write for a vector $V_M$ or arbitrary density weight
 \be
   \big({\cal L}^{\nabla}_{(\Lambda,\widetilde{\Sigma})}   - {\cal L}^{}_{(\Lambda,\Sigma)}\big) V_M
  \ = \ -{\cal T}_{MN,KL}\,V^N\Lambda^{KL}\;, 
 \ee 
in terms of (\ref{FULLTorsion}).  
This follows by a direct computation. 
Useful intermediate results are (recalling that $\Lambda$ has weight $\lambda=1$) 
 \be
 \begin{split}
  \nabla_{MN}\Lambda^{MN} \ &= \ \partial_{MN}\Lambda^{MN} +\Gamma_{M}{}^{K}{}_{,NK}\Lambda^{MN}\;, \\
  K_{MN}^{\nabla}(\Lambda,\Sigma) \ &= \ K_{MN}(\Lambda, \Sigma) 
  +\Lambda^{KL}\big(\Gamma_{MN,KL}+\Gamma_{KL,MN}-{\cal T}_{MNKL}\big)
  -2\,{\cal T}_{[M}{}^{K}\Lambda_{|K|N]}\;. 
 \end{split}  
 \ee

With these relations we can relate the general variation (\ref{covCSvar}) of the Chern-Simons term 
to its `covariant variation' as used in \cite{Baguet:2016jph}. Indeed, one quickly sees, 
upon adding and subtracting 
connection terms, that  
 \be
 \begin{split}
  \delta S_{\rm CS} \ = \  \int {\rm d}^3x\,{\rm d}^NY\,\varepsilon^{\mu\nu\rho}\Big(&\, \Delta {\cal B}_{\mu MN} \,{\cal F}_{\nu\rho}{}^{MN}
   +\delta {\cal A}_{\mu}{}^{MN}\big(\tilde{{\cal G}}_{\nu\rho\, MN} + \nabla_{MK}{\cal  F}_{\nu\rho\, N}{}^{K} 
\big)\Big)\;, 
 \end{split} 
 \ee  
where we introduced 
 \be
  \begin{split}
   \Delta {\cal B}_{\mu MN} \ &\equiv \ \delta {\cal B}_{\mu MN} 
   - \frac{1}{4}\,\Gamma_{MN,KL}\,\delta {\cal A}_{\mu}{}^{KL}\;, \\
   \tilde{{\cal G}}_{\mu\nu \,MN} \ &\equiv \ {\cal G}_{\mu\nu \, MN} - \frac{1}{4}\,\Gamma_{MN,KL}\,{\cal F}_{\mu\nu}{}^{KL}\;. 
  \end{split}
 \ee  
Let us also note with (\ref{conntransform}) that $\tilde{{\cal G}}$ transforms as  
 \be
  \delta_{\Upsilon}\tilde{{\cal G}}_{\mu\nu\,MN} \ = \ {\cal L}_{\Upsilon}\tilde{{\cal G}}_{\mu\nu\,MN}\;,  
 \ee
i.e., it transforms covariantly in the more conventional sense of covariance.

\section{Construction of \texorpdfstring{${\rm O}(p,q)$}{O(p,q)} enhanced DFT}

Having set up the formalism we can now construct the 
enhanced DFT invariant under ${\rm O}(p,q)$ generalized diffeomorphisms.
The field content of the ${\rm O}(p,q)$ enhanced DFT is given by
the gauge fields (\ref{DoubledVEctor}) together with
an external $3\times 3$ metric $g_{\mu\nu}$ (or vielbein $e_\mu{}^a$),
and an internal ${\rm O}(p,q)$ valued metric ${\cal M}_{MN}$.

\subsection{Building blocks of the DFT action}

The field equations of ${\rm SO}(p,q)$ enhanced DFT
are most compactly derived from a Lagrangian whose different terms
are of the form generic for exceptional field theory with 
three external dimensions~\cite{Hohm:2013jma,Hohm:2014fxa}
\bea
{\cal L} &=& 
{\cal L}_{\rm EH} + k_1\, {\cal L}_{\rm kin} +  k_2\,{\cal L}_{\rm CS} +k_3\, {\cal L}_{\rm pot}
\;,
\label{action0}
\eea
each term being separately invariant under generalized internal diffeomorphisms. 
The modified Einstein-Hilbert term and the scalar kinetic term have the following form
\bea
\mL_{\rm EH}&=&  \sqrt{-g}\,e_a{}^\m e_b{}^\n \left(
{R}_{\m\n}{}^{ab}+ {\cal F}_{\m\n}{}^{M N}e^{a\,\r}\dt_{MN}e_{\r}{}^{b}
\right)
~\equiv~  \sqrt{-g}\,\hat{R}\nonumber\;,\\
\mL_{\rm kin}&=&\fr{1 }{16}\,   \sqrt{-g}\,\,g^{\m\n}\,D_{\m}\cM^{MN} D_\n \cM_{MN},
\label{LEH}
\eea
with the covariant derivatives (\ref{genericcovDER}) and the Riemann
tensor ${R}_{\m\n}{}^{ab}$ computed from the external vielbein $e_\mu{}^a$
with derivatives covariantized under internal diffeomorphisms under which
$e_\mu{}^a$ transforms as a scalar density (of weight $\lambda=1$).
By construction, both these terms are invariant under generalized internal
diffeomorphisms with the second term in $\hat{R}$ moreover ensuring invariance
under local ${\rm SO}(1,2)$ Lorentz transformations.\footnote{
Note the absence of the factor of $1/2$  with respect to the expression 
in \cite{Musaev:2015ces} that is due to our different sum conventions for
sums over pairs of antisymmetric indices.} 

The Chern-Simons term in (\ref{action0}) is given by the standard non-abelian form
(\ref{SCS}) based on the gauge invariant inner product (\ref{prod}) on the gauge algebra
of internal diffeomorphisms. For concreteness, we spell out its explicit form
\bea
{\cal L}_{\rm CS} &=&
\varepsilon^{\mu\nu\rho}\,\Big( 
{F}_{\mu\nu}{}^{MN} {\cal B}_{\rho\,MN}
+ \partial_\mu {\cal A}_{\nu\,N}{}^K \partial_{K M} {\cal A}_\rho{}^{MN}
-\frac23\,
\partial_{MN}\partial_{KL} {\cal A}_{\mu}{}^{KP} {\cal A}_\nu{}^{MN} {\cal A}_{\rho\,P}{}^{L}
\nonumber\\
&&{}
\qquad\qquad
+\frac23\, {\cal A}_{\mu}{}^{LN} \partial_{MN} {\cal A}_\nu{}^{M}{}_{P} \partial_{KL} {\cal A}_\rho{}^{PK}
-\frac43\,{\cal A}_{\mu}{}^{LN} \partial_{MP} {\cal A}_\nu{}^{M}{}_{N} \partial_{KL} {\cal A}_\rho{}^{PK}\Big) 
\;,\quad
\label{CS}
\eea
with its variation given by (\ref{covCSvar}).

The last term in (\ref{action0}) is referred to as the {\em potential term} (from a three-dimensional point of view)
as it does not carry any external derivatives $\partial_\mu$, but is bilinear in the internal currents
\bea
{\cal J}_{MN,KL} &\equiv&{\cal M}^{PQ}\eta_{QK}  \partial_{MN} {\cal M}_{LP}\,
\;, \nonumber\\
({\cal J}_{MN})_\mu{}^\nu &\equiv& g^{\nu\rho}\,\partial_{MN} g_{\mu\rho}
\;,
\label{JJint}
\eea
such that the full expression  is invariant under
generalized internal diffeomorphisms up to total derivatives.
It is useful to note the non-covariant transformation behavior of the currents (\ref{JJint})
\bea
{\cal L}_{(\Lambda,\Sigma)} {\cal J}_{MN,KL} &=&
\delta^{\rm cov} {\cal J}_{MN,KL} +
\left({\cal M}_{P[K}{\cal M}_{L]Q} - {\eta}_{P[K}{\eta}_{L]Q}\right) 
\partial_{MN} K^{PQ}
\;,
\nonumber\\
{\cal L}_{(\Lambda,\Sigma)} {\cal J}_{\mu}{}^{\nu}&=&
\delta^{\rm cov}{\cal J}_{\mu}{}^{\nu} +
2\,\partial_{MN}\partial_{KL}\Lambda^{KL}\,\delta_{\mu}{}^{\nu}
\;,
\eea
with $K^{PQ}$ from (\ref{KDEFF}).
It is then straightforward to verify by explicit calculation that the following combination of currents
\bea
V &\equiv&
-\frac18\,
{\cal M}^{KP}{\cal M}^{LQ}\,
\partial_{KL} {\cal M}_{MN}\,
\partial_{PQ} {\cal M}^{MN}
-\frac12\,\partial_{MK} {\cal M}^{NP}
\partial_{NL} {\cal M}^{MQ}\,{\cal M}^{KL}{\cal M}_{PQ} 
 \nonumber\\
 &&{}
-\frac14\,
\partial_{MN} {\cal M}^{PK}\,
\partial_{KL} {\cal M}^{QM}\,{\cal M}_{P}{}^L{\cal M}_{Q}{}^N
\nonumber
+2\,\partial_{MK} {\cal M}^{NK}\,\partial_{NL} {\cal M}^{ML}
 \nonumber\\
 &&{}
-g^{-1}\partial_{MN} g\,\partial_{KL}{\cal M}^{MK}{\cal M}^{NL}
 -\frac{1}{4}  \,{\cal M}^{MK}{\cal M}^{NL}\,g^{-2}\partial_{MN}g\,\partial_{KL}g
\nonumber\\
 &&{}
  -\frac{1}{4}\,{\cal M}^{MK}{\cal M}^{NL}\,\partial_{MN}g^{\mu\nu}\partial_{KL} g_{\mu\nu}
  \;,
  \label{full_potential}
\eea
is such that ${\cal L}_{\rm pot}\equiv -\sqrt{-g}\,V$ is indeed invariant under
generalized internal diffeomorphisms up to total derivatives.

The Lagrangian (\ref{action0}) thus is (term by term)
invariant under internal generalized diffeomorphisms
up to total derivatives $\sqrt{-g}^{-1} \partial_\mu \left(\sqrt{-g} I^\mu\right)$\,.
It remains to fix the relative coefficients $k_1$, $k_2$, $k_3$.
This will be a consequence of the invariance under external diffeomorphisms.

\subsection{External diffeomorphisms}

The full Lagrangian (\ref{action0}) should also be invariant under a suitable definition
of external diffeomorphisms with parameter $\xi^\mu(x,Y)$. 
This fixes all remaining constants in the Lagrangian. The calculation closely 
follows the analogous cases of maximal E$_{8(8)}$ ExFT \cite{Hohm:2014fxa}
and minimal SL(2) ExFT \cite{Hohm:2013jma}, such that here we only briefly sketch
the pertinent cancellations in order to determine the constants $k_1$, $k_2$, $k_3$\,.
For the external dreibein field and the scalar matrix external diffeomorphisms
take the usual form
\begin{equation}
\label{ext_diff}
 \begin{aligned}
  \d e_\m{}^a &=\x^\m {D}_\n e_\m{}^a+{D}_\m \x^\n e_\n{}^a\;,\\
  \d \cM_{MN} &= \x^\m {D}_\m \cM_{MN},
 \end{aligned}
\end{equation}
of properly covariantized three-dimensional diffeomorphisms. 
For the gauge fields, we start from an ansatz following  \cite{Hohm:2013jma,Hohm:2014fxa} 
\begin{equation}
\label{ext_diff2}
 \begin{aligned}
     \d_\x^{(0)} \cA_\m{}^{M N} &= \x^\n F_{\n\m}{}^{MN}+
     g_{\m\n}\cM^{MK}\cM^{NL}\dt_{KL}\x^\n,\\
     \d_\x^{(0)}\cB_{\m\, M N} &= \x^\n G_{\n\m\, MN}+
     \b_1 g_{\m\n} {\cal J}_{MN}{}^{KL}\dt_{KL}\x^\n+\b_2\,\sqrt{-g}\, \ve_{\m\n\l}g^{\l\r}\mD^\n(g_{\r\s}\dt_{MN}\x^\s),
 \end{aligned}
\end{equation}
which reduces to standard three-dimensional diffeomorphisms in case the parameter $\xi^\mu$ does
not depend on the internal coordinates. The coefficients $\b_1, \b_2$ will be fixed in the following.

In what follows it proves useful to have the explicit form of variation of the full 
Lagrangian with respect to a variation of the gauge fields which we put in the form
\bea
\label{var_AB}
\d_{(\cA,B)}\, \mL&=&\ve^{\m\n\r}\,
\Big({\cal E}_{\m\n}^{(\cA)MN}\,\d\cB_{\r\, MN} 
+{\cal E}_{\m\n\, MN}^{(\cB)}\,\d \cA_\r{}^{MN}\Big)
\;,
\eea
with the coefficients
\bea
{\cal E}_{\m\n}^{(\cA)MN}&=& k_2\,   F_{\m\n}{}^{MN} 
-\fr12\,\sqrt{-g}\,  k_1 \, \ve_{\m\n\s}\,j^{\s\, MN}\;,\nonumber\\
{\cal E}_{\m\n\, MN}^{(\cB)}&=&  k_2\,    G_{\m\n MN} +
\sqrt{-g}\,\ve_{\m\n\s}\,\hat{J}^\s{}_{MN} -\fr{1}{16} \,k_1\,  \sqrt{-g}\,\ve_{\m\n\s}j^{\s K}{}_{L}{\cal J}_{MN}{}^L{}_K\nonumber\\
&&{} + \dt_{MK}{\cal E}_{\m\n}^{(\cA)}{}_{N}{}^K
\;.
\label{EAEB}
\eea
Here, the internal current ${\cal J}_{MN}{}^L{}_K$ has been defined in (\ref{JJint}),
the external currents are defined as
\begin{equation}
\begin{aligned}
j_\m{}^{MN}&=\h_{KL}\,\cM^{K[M}\mD_\m\cM^{N]L},  \\
\hat{J}^\m{}_{MN}&=-2\,e^\mu{}_a\, e^\n{}_b\, \Big( \dt_{MN} \w_\n{}^{ab} -D_\n(e^{\r [a}\dt_{MN} e_\r{}^{b]}) \Big)\;,
\end{aligned}
\end{equation}
and related to the sources from the Einstein-Hilbert and the kinetic scalar term, respectively,
c.f.\ (\ref{varEH}) below.
Note that the first equation of (\ref{EAEB}) does not appear as a full field equation of the theory, 
since the field ${\cal B}_{\mu\,MN}$
w.r.t.\ which we vary in (\ref{var_AB}) is constrained according to (\ref{sectionB}).

With the variation (\ref{ext_diff}), (\ref{ext_diff2}) and the general variation (\ref{covCSvar}) of the Chern-Simons term,
we find that under external diffeomorphisms this term transforms non-trivially as
\bea
\label{var_cs}
\d_\x^{(0)}\mL_{\rm CS}&=& 
 \ve^{\m\n\r}
\left(
\cM^{MP}\cM^{NQ}\dt_{MK} F_{\m\n N}{}^{K}g_{\r\s}\dt_{PQ}\x^\s
+   G_{\m\n PQ}\cM^{PK}\cM^{QL}g_{\r\s}\dt_{KL}\x^\s \right)
\nonumber\\
&&{}+\b_1\, \ve^{\m\n\r} F_{\m\n}{}^{PQ}{\cal J}_{PQ}{}^{KL}g_{\r\s}\dt_{KL}\x^\s
-2\, \b_2\, \sqrt{-g}  \, F^{\m\n MN}\mD_\m(\dt_{MN}\x^\r g_{\r\n})\nonumber\\
&&{}-\fr{1 }{2} \,  \ve^{\m\n\r}\dt_{MK}\x^{\s}   F_{\s\r}{}^{MN} F_{\m\n N}{}^{K}\;,
\eea
up to total derivatives. Using (\ref{EAEB}), the last term here can be written as
\bea
-\fr{1 }{2}\, \ve^{\m\n\r}\,\dt_{MK}\x^{\s}   F_{\s\r}{}^{MN} 
 F_{\m\n N}{}^{K}&=&
-\fr{1}{2\,k_2^2 }\,  \ve^{\m\n\r}\dt_{MK}\x^\s {\cal E}_{\s\r}^{(\cA)MN}{\cal E}_{\m\n }^{(\cA)}{}_N{}^K\\
&&{}
-\left(\frac{k_1}{k_2^2}\, \sqrt{-g}  \, F_{\n\rho}{}^{MN} 
-\fr{k_1 ^2}{4 \,k_2^2 } \, \ve_{\m\n\r}\,j^{\m\, MN}\right) j^\n{}_N{}^K \dt_{MK}\x^\r\;.
\nonumber
\eea
Next we proceed with variation of the Einstein-Hilbert term. With its variation
under a general variation of the gauge field $\cA_\m{}^{MN}$ given by
\begin{equation}
\d_\cA \mL_{\rm EH}= \hat{J}^\m{}_{MN}\,\d \cA_\m{}^{MN}\;,
\label{varEH}
\end{equation}
the full diffeomorphism variation of the covariantized EH term becomes
up to total derivatives
\begin{equation}
\d_\x^{(0)}(\sqrt{-g} \hat{R})=\sqrt{-g} \, F^{\m\n MN}\mD_\m(\dt_{MN}\x^\r g_{\r\n})+
\sqrt{-g}\, \cM^{MK}\cM^{NL}\hat{J}_{\m\,KL}\dt_{MN}\x^\m\;.
\end{equation}
The first term in this variation has been computed in \cite{Hohm:2013jma}
and cancels the corresponding term in the variation of the Chern-Simons term 
if we choose $\b_2 =1/(2\,k_2)$.

Also the variation of the scalar kinetic term follows \cite{Hohm:2013jma}. We find
\bea
\label{var_sc}
\delta{\cal L}_{\rm kin}&=& \delta^{\rm cov}\,\delta{\cal L}_{\rm kin}
+\sqrt{-g}\,j^\m{}^{R}{}_{Q}\dt_{PR}\big(g_{\m\r}\cM^{PK}\cM^{QL}\dt_{KL}\x^\r\big) \nonumber\\
&&{}-\fr 18\,\sqrt{-g}\,  g_{\m\n}\cM^{PK}\cM^{QL}\dt_{PQ}\x^\n {\cal J}_{KL}{}^{MN}j_{\n MN}-\,\sqrt{-g}\,
 F_{\m\n}{}^{KQ}\dt_{KL}\x^\n j^{\m L}{}_{Q}
\nonumber\\
&&{}+\sqrt{-g}\, \b_1e {\cal J}_{KL}{}^{PQ}j_\n{}^{KL}\dt_{PQ}\x^\n +\frac{1}{2k_2}\,
 \ve^{\m\n\r}j_\m{}^{KL}\mD_\n\big(g_{\r\s}\dt_{KL}\x^\s\big)
\;.
\eea
Upon integrating by parts the derivative $\mD_\n$ in the last term above it can be rewritten in the following form
\begin{equation}
\begin{aligned}
&-\frac{1}{2k_2}\,
 D_\n\big( \ve^{\m\n\r}j_\m{}^{KL} \big) g_{\r\s}\dt_{KL}\x^\s
 =\\
=&\ \frac{1}{2k_2}\,\ve^{\m\n\r}j_\n{}^K{}_N j_\m{}^{NL}g_{\r\s}\dt_{KL}\x^\s+\frac{1}{4k_2}\, \ve^{\m\n\r}\cM^{K}{}_{M}\mL_{( F_{\n\m}, G_{\n\m})}\cM^{ML}
g_{\r\s}\dt_{KL}\x^\s\\
=&\ \frac{1}{2k_2}\,\ve^{\m\n\r}j_\n{}^K{}_N j_\m{}^{NL}g_{\r\s}\dt_{KL}\x^\s+\frac{1}{4k_2}\ve^{\m\n\r} F_{\n\m}{}^{PQ}
\,{\cal J}_{PQ}{}^{KL}g_{\r\s}\dt_{KL}\x^\s\\
&+\frac{1}{k_2}\,\ve^{\m\n\r}\,\cM^{MK}\cM^{NL} \dt_{M}{}^P F_{\n\m\,NP}{}\,g_{\r\s}\,\dt_{KL}\x^\s\\
&+\frac{1}{k_2}\, \ve^{\m\n\r} G_{\n\m PQ}\cM^{PK}\,\cM^{QL}g_{\r\s}\dt_{KL}\x^\s\;.
\end{aligned}
\end{equation}
Here, the terms linear in the field strengths cancel the corresponding terms from the variation of the Chern-Simons term \eqref{var_cs} if the following holds true
\begin{equation}
k_1 = k_2^2 \;,\qquad \beta_1 = \frac{1}{4}
\;.
\end{equation}

The remaining contributions coming from the Einstein-Hilbert, the scalar kinetic and the Chern-Simons terms can be collected in the following expression 
\begin{equation}
\begin{aligned}
&\d_\x^{(0)}\Big(\mL_{\rm EH}+k_1\,\mL_{\rm kin}+k_2\,\mL_{\rm CS}\Big)=\\
=&\   \sqrt{-g}\, \cM^{MK}\cM^{NL}\hat{J}_{\m\,KL}\dt_{MN}\x^\m
+k_1  \,\sqrt{-g}\,j^\m{}^{L}{}_{Q}\dt_{PL}\big(g_{\m\r}\cM^{PK}\cM^{QL}\dt_{KL}\x^\r\big) \\
&-\fr 18k_1  \,\sqrt{-g}\, g_{\m\n}\cM^{PK}\cM^{QL}\dt_{PQ}\x^\n j_{KL}{}^{MN}j_{\n MN}
\\
&+k_1 \b_1\,\sqrt{-g}\, {\cal J}_{KL}{}^{PQ}j_\m{}^{KL}\dt_{PQ}\x^\m
-\fr{1}{2\,k_2 }\ve^{\m\n\r}\dt_{MK}\x^\s {\cal E}_{\s\r}^{(\cA)MN}{\cal E}_{\m\n }^{(\cA)}{}_N{}^K\\
&+\fr{k_1^2}{4 \,k_2}(-g)\,\varepsilon_{\m\n\r}j^{\m\, MN}j^\n{}_N{}^K \dt_{MK}\x^\r+ \frac{k_1}{2\,k_2}\,\ve^{\m\n\r}j_\n{}^K{}_N j_\m{}^{NL}g_{\r\s}\dt_{KL}\x^\s
\end{aligned}
\label{varRest}
\end{equation}
Terms in the last line cannot be cancelled by any contribution coming from the scalar potential and hence must cancel each other, for which we must choose $k_1 =2$\,.

To see the cancellations coming from the variation of the scalar potential let us look only at the relevant terms inside variation of the potential \eqref{full_potential}. First it is useful to write first variations of the scalar current ${\cal J}_{MN}{}^{KL}$ and of the derivative $\dt_{MN}g_{\m\n}$ that read
\begin{equation}
\begin{aligned}
\d_\x\big({\cal J}_{MN}{}^K{}_L\big)&= 
\x^\m \mD_\m\big({\cal J}_{MN}{}^K{}_L\big) +\dt_{MN}\x^\m\, j_\m{}^K{}_L,\\
\d_\x\big(\dt_{MN}g_{\m\n}\big)&=\mL_\x \big(\dt_{MN}g_{\m\n}\big)
+\dt_{MN}\,\x^\r \mD_\r g_{\m\n}+2\big(\dt_{MN}\mD_{(\m}\x^\r\big)g_{\n)\r}.
\end{aligned}
\end{equation}
The first term in each line is a covariant variation, while the remaining parts give the non-covariant variation of the scalar potential.
Since the full cancellations work precisely like in the $E_{8(8)}$ theory \cite{Hohm:2014fxa} there is no need to repeat the full derivation here. Let us check the most indicative terms to fix the coefficients and to check the consistency. For that we consider the following contribution from the non-covariant variation of ${\cal L}_{\rm pot}$
\begin{equation}
\begin{aligned}
- k_3\,\d \left( \sqrt{-g} \,V\right)&=
- k_3\,\d^{\rm cov} \left( \sqrt{-g} \,V\right)
-\fr{k_3}{2}  \sqrt{-g}\, \dt_{KL}\x^\m j_\m{}^{MN}{\cal J}_{MN}{}^{KL}+\ldots,
\end{aligned}
\end{equation} 
whose cancellation against the corresponding term in (\ref{varRest}) 
forces us to set  $k_3=2k_1 \b_1=1$\,. 
We have now fixed all the unknown coefficients in (\ref{action0}) and (\ref{ext_diff2})
\bea
k_1=2\;,\quad
k_2=\sqrt{2}\;,\quad
k_3=1\;,\qquad
\beta_1=\frac14\;,\quad
\beta_2=\frac1{2\sqrt{2}}
\;.
\eea
After these numerical values
ensure all the above cancellations to take place 
we are finally left with  the  following variation of the full Lagrangian 
\begin{equation}
\label{last_piece}
\begin{aligned}
\d_\x^{(0)}\Big(\mL_{\rm EH}+2\,\mL_{\rm kin}+\sqrt{2}\,\mL_{\rm CS}+{\cal L}_{\rm pot}\Big)
&=-\fr{1}{2\sqrt{2} }\varepsilon^{\m\n\r}\dt_{MK}\x^\s {\cal E}_{\s\r}^{(\cA)MN}{\cal E}_{\m\n }^{(\cA)}{}_N{}^K\\
&=\fr{1}{\sqrt{2}}\varepsilon^{\m\n\r}\x^\s \dt_{MK}{\cal E}_{\s\r}^{(\cA)MN}{\cal E}_{\m\n }^{(\cA)}{}_N{}^K
\;,
\end{aligned}
\end{equation}
up to total derivatives.
To get rid of this remnant, we perform the same trick as in \cite{Hohm:2014fxa} and define the full diffeomorphism transformation of the gauge fields as the following deformation of the initial ansatz (\ref{ext_diff2})
\begin{equation}
\begin{aligned}
\d_\x \cA_\m{}^{MN}&=\d^{(0)}_\x \cA_\m{}^{MN}+\fr{1}{\sqrt{2}  } \x^\n {\cal E}_{\m\n}^{(\cA) MN} \\
\d_\x\cB_{\m\, MN}&=\d^{(0)}_\x\cB_{\m\, MN}+\fr{1}{\sqrt{2} } \x^\n \Big( {\cal E}_{\m\n}^{(\cB)}{}_{MN} -\fr{1}{8}f_{MN,KL}{}^{PQ}\dt_{PQ}{\cal E}_{\m\n}^{(\cA)KL}\Big)\\
&=\d^{(0)}_\x\cB_{\m\, MN}+ \fr{1}{\sqrt{2}  }\x^\n \Big( {\cal E}_{\m\n}^{(\cB)}{}_{MN} - \dt_{K[M}{\cal E}_{\m\n}^{(\cA) K}{}_{N]}\Big).
\end{aligned}
\end{equation}
Indeed, according to \eqref{var_AB} and the above discussion, the variations $\d^{(0)}_\x$ provide  the contribution \eqref{last_piece} which cancels against the term coming from the $\dt {\cal E}^{(\cA)}$ in the second line. The new contributions of the form ${\cal E}^{(\cA)}\cdot {\cal E}^{(\cB)}$ cancel each other as they form an expression totally antisymmetric in four space-time indices. The mutual factor in the brackets of the second and the last line above was chosen in such a way as to keep $\d_\x\cB_{\m\, MN}$ satisfying the same section constraints as the field ${\cal B}_{\m\, MN}$ does. 

Hence, the full diffeomorphism transformations leaving the theory invariant can be collected as follows
\begin{equation}
\begin{aligned}
\d e^a_\m &=\x^\m \mc{D}_\n e^a_\m+\mc{D}_\m \x^\n e^a_\n\;,\qquad
  \d \cM_{MN} = \x^\m \mc{D}_\m \cM_{MN}\;,\\
\d_\x \cA_\m{}^{M N} &=  -\sqrt{-g}\,\x^\n   \ve_{\m\n\s}j^{\s MN}
+g_{\m\n}\,\cM^{MK}\cM^{NL}\dt_{KL}\x^\n\;,\\
\d_\x \cB_{\m\, M N} &= 
\sqrt{-g}\,\ve_{\m\n\r}\Big(\fr{1}{2\sqrt{2}} g^{\l\r}\mD^\n(g_{\l\s}\dt_{MN}\x^\s)  
     + \x^\n \hat{J}^\r{}_{MN} -\fr{1}{8}\x^\n j^{\r}{}_{KL}{\cal J}_{MN}{}^{KL} \,\Big)\\
      &\quad +\fr{1}{2} \,g_{\m\n} \,{\cal J}_{MN}{}^{KL}\dt_{KL}\x^\n\;,
\end{aligned}
\label{diff_ex}
\end{equation}
that have precisely the same form as the ones in \cite{Hohm:2014fxa} as expected. The final Lagrangian then becomes
\begin{equation}
\label{action_fixed}
\begin{aligned}
{\cal L} &= 
\mL_{\rm EH}+2\,\mL_{\rm kin}+\sqrt{2}\,\mL_{\rm CS}+{\cal L}_{\rm pot}
\;,
\end{aligned}
\end{equation}
with all relative coefficients now fixed by invariance under external diffeomorphisms (\ref{diff_ex}).

\subsection{Solutions of the section constraint}
\label{subsec:solutions}

Let us now discuss the explicit solutions of the section constraint (\ref{section}).
We will identify two inequivalent solutions that essentially correspond to the 
embedding of $D=6$ non-chiral and chiral theories, respectively.

For the first solution, we start from the theory based on ${\rm O}(d+1,d+1+n)$
and consider its decomposition under ${\rm GL}(d)$ embedded as
\bea
{\rm O}(d+1,d+1+n) &\supset& {\rm O}(d,d) ~\supset~ {\rm GL}(d)\;,
\eea
with fundamental vectors breaking into
\bea
\left\{V^{M}\right\}&\longrightarrow&
\left\{V^{i}, V^0, V_{i}, V_0,  \tilde{V}^{p}\right\}\;,\quad
i=1, \dots, d\;,\quad p=1, \dots, n
\;,
\eea
and a Cartan-Killing form
\bea
\eta_{MN} &=& 
\begin{pmatrix}
0_{d\times d} & 0 & \delta_i{}^j & 0& 0_{d\times n}\\
0&0&0&1&0\\
\delta^i{}_j &  0& 0_{d\times d}& 0& 0_{d\times n}\\
0&1&0&0&0\\
0_{n\times d} &0& 0_{n\times d} &0& \mathbb{I}_{n\times n}
\end{pmatrix}
\;.
\eea
It is then straightforward to see that 
restricting all fields to depend exclusively on $d$ coordinates $y^i$ defined as
\bea
\left\{y^i \equiv Y^{i0}\right\}
\;,\qquad
\Phi(x^\mu, Y^{MN}) ~=~ \Phi(x^\mu, y^i)\;,
\label{sec_sol_1}
\eea
constitutes a solution to (\ref{section}).\footnote{
More elaborately, we
could in a first step have broken down ${\rm O}(d+1,d+1+n)$ to ${\rm O}(d,d)$ and selected
coordinates $\{Y^I\}\equiv\{  Y^{i0}, Y_{i0} \}$, such that the section constraints (\ref{section})
reduce to $\eta^{IJ}\,\partial_I\otimes\partial_J=0$ and reproduce the 
structures of standard double field theory. In a second step, this remaining section constraint
is then solved by (\ref{sec_sol_1}).
}
Upon evaluating the above constructed theory for this solution of the section constraint, 
it reproduces the field equations of the bosonic string in $d+3$ dimensions,
coupled to $n$ abelian vectors,
i.e., for $n=16$ the field equations of the heterotic string truncated to the 
Cartan subalgebra of the full gauge group.

An alternative solution to the section constraints (\ref{section}) 
is found by starting from the theory based on ${\rm O}(3+n_{\rm L},3+n_{\rm R})$
and decomposing it under a ${\rm GL}(3)$ embedded as
\bea
{\rm O}(3+n_{\rm L},3+n_{\rm R})&\supset& {\rm O}(3,3)~\supset~
{\rm GL}(3)\;,
\eea
with fundamental vectors breaking into
\bea
\left\{V^{M}\right\}&\longrightarrow&
\left\{V^{i},  V_{i},   \tilde{V}^{p},   \bar{V}^{q}\right\}\;,\quad
\nonumber\\
&&{}
i=1, \dots, 3\;,\quad p=1, \dots, n_{\rm L}\;,\quad q=1, \dots, n_{\rm R}
\;,
\eea
and a Cartan-Killing form
\bea
\eta_{MN} &=& 
\begin{pmatrix}
0_{d\times d}  & \delta_i{}^j & 0_{d\times n_{\rm L}}& 0_{d\times n_{\rm R}}\\
\delta^i{}_j &  0_{d\times d}&  0_{d\times n_{\rm L}}& 0_{d\times n_{\rm R}}\\
0_{n\times d} & 0_{n\times d} & \mathbb{I}_{n_{\rm L}\times n_{\rm L}}  & 0_{n_{\rm L} \times n_{\rm R}}\\
0_{n\times d} & 0_{n\times d} &0_{n_{\rm R} \times n_{\rm L}} & -\mathbb{I}_{n_{\rm R}\times n_{\rm R}} 
\end{pmatrix}
\;.
\eea
Restricting all fields to depend exclusively on coordinates $\tilde{y}_i$ defined as
\bea
\left\{\tilde{y}_i\equiv \varepsilon_{ijk} Y^{jk}\right\}
\;,\qquad
\Phi(x^\mu, Y^{MN}) ~=~ \Phi(x^\mu, \tilde{y}_i)\;,
\label{sec_sol_2}
\eea
again constitutes a solution to (\ref{section}).
In this case, the above constructed theory reproduces the
field equations of $D=6$ gravity, coupled to $n_{\rm L}$ selfdual
and $n_{\rm R}$ anti-selfdual antisymmetric two-form tensors,
as well as to $n_{\rm L}\cdot n_{\rm R}$ scalar fields.\footnote{
In particular, the special case $n_{\rm L}=n_{\rm R}=0$ corresponds to pure $D=6$ gravity
with ${\rm SO}(3,3) \sim {\rm SL}(4)$ encoding the Ehlers symmetry group
upon reduction to three dimensions. The gauge structure and section constraints
in this case have also been considered in \cite{Strickland-Constable:2013xta,Cederwall:2015ica}.
}
Indeed, it follows from inspection that fields depending on the full set of coordinates $\{\tilde{y}^i\}$ cannot 
depend on any further internal coordinate without violating the section constraints (\ref{section}).
The resulting theory cannot be lifted beyond six dimensions which is the case for the chiral theories
coupling (anti-)selfdual tensor fields.\footnote{
{Similarly, the section constraints in exceptional field theory in general
admit two inequivalent solutions corresponding to a higher-dimensional IIA and IIB origin~\cite{Hohm:2013pua}.
Specifically, the two solutions (\ref{sec_sol_1}), (\ref{sec_sol_2}) are based on different embeddings of ${\rm GL}(3)$ into
${\rm SO}(3,3)$, in analogy to the two inequivalent solutions in ${\rm SL}(5)$ exceptional field theory~\cite{Blair:2013gqa}.
}
}

Comparing the two solutions (\ref{sec_sol_1}), (\ref{sec_sol_2}) it is obvious that
for $d\le2$ the coordinates (\ref{sec_sol_1}) can be considered as
a subset of (\ref{sec_sol_2}). Indeed, in this case
 the $D\le5$ theories described by (\ref{sec_sol_1}) are obtained by dimensional
 reduction (and possible truncation) from the $D=6$ theories described by (\ref{sec_sol_2}).
The two solutions thus are not independent. For $d>3$ on the other hand,
the different choices of coordinates are inequivalent (as discussed, the set of coordinates 
(\ref{sec_sol_2}) cannot be extended without violating the section constraints (\ref{section}),
thus never be equivalent to the $d>3$ coordinates (\ref{sec_sol_1})) ---
and so are the resulting higher-dimensional theories.
An interesting case is the theory with $d=3$, $n=0$ (i.e.\ $n_{\rm L}=n_{\rm R}=1$),
built on the group ${\rm O}(4,4)$. In this case, the two choices of coordinates
(\ref{sec_sol_1}) and (\ref{sec_sol_2}) can be shown to be related by an outer
automorphism (a triality rotation) of ${\rm SO}(4,4)$, they hence describe equivalent theories.
Indeed, the $D=6$ theory from (\ref{sec_sol_2}) coupling gravity to one selfdual tensor, one anti-selfdual tensor,
and a scalar field, is precisely the bosonic string described by (\ref{sec_sol_1}).
We will come back to this equivalence later when discussing Scherk-Schwarz reductions
and consistent truncations.

Let us finally discuss two important series of theories, based on the groups
${\rm O}(4,n)$ and ${\rm O}(8,n)$, respectively. These theories can be supersymmetrized
upon adding fermionic fields into half-maximal and quarter-maximal field theories, respectively.
According to the above discussion, the ${\rm O}(4,4)$ theory has a unique solution of the section
constraint which describes the embedding of the $D=6$, ${\cal N}=(1,0)$ supergravity coupled to one tensor multiplet,
such that its full field content and couplings are non-chiral. The theories built from ${\rm O}(4,4+n)$, $n>0$, 
on the other
hand admit two inequivalent solutions (\ref{sec_sol_1}), (\ref{sec_sol_2}) of the section constraint,
describing the coupling of ${\cal N}=(1,0)$ vector multiplets and chiral tensor multiplets, 
respectively, to this $D=6$ supergravity.
The ${\rm O}(8,n)$ theories can be supersymmetrized
into half-maximal field theories. For these theories, the solution (\ref{sec_sol_1}) of the section constraint, 
describes the embedding of $D=(2+n)$ half-maximal supergravity for $n\le8$ and of $D=10$, ${\cal N}=1$
supergravity with $n-8$ vector multiplets for $n\ge8$, respectively.
The solution (\ref{sec_sol_2}) on the other hand describes the embedding of $D=6$, ${\cal N}=(2,0)$ 
chiral supergravity coupled to $n-3$ tensor multiplets. In accordance with the above counting,
every one of these multiplets combines a selfdual tensor with five scalar fields
while the ${\cal N}=(2,0)$ supergravity multiplet carries five anti-selfdual tensors.

Table~\ref{tab:sol} summarizes the embedding of the various higher-dimensional theories.
For completeness, let us mention that the theory based on the group ${\rm O}(2,1)$ constructed 
in~\cite{Hohm:2013jma} which describes pure $D=4$ gravity with the Ehlers group made manifest, does not seem to 
fit in the present construction. This is seen from the fact the section constraints (\ref{section}) for ${\rm O}(2,1)$
do not admit any solution whereas the construction of ~\cite{Hohm:2013jma} is based on a weaker version
of the section constraints (suppressing only the ${\bf 1}\oplus {\bf 3}$ in ${\bf 3}\otimes{\bf 3}$) which allows
for a one-dimensional solution.

\begin{table}[htb]
\begin{tabular}{|rccll|l|}
\hline
 ${\rm O}(d+1, d+1+n)$ &$ \hookleftarrow$ & $ {\rm O}(d,d) $ & $ \hookleftarrow $ & $ {\rm GL}(d)$ & $D=d+3$  bosonic string,  \\
    $Y^{MN}$ & $ \longrightarrow $ & $ (Y^{i0}, Y_{i0}) $ & $ \longrightarrow  $ & $ y^i=Y^{i0}$ & with \eqref{sec_sol_1} and $n_{\rm V}=n$ \\
   \hline 
  ${\rm O}(3+n_L, 3+n_R)$ &$ \hookleftarrow$ & $ {\rm O}(3,3) $ & $ \hookleftarrow $ & $ {\rm GL}(3)$ &  $D=6$ gravity,  \\
    $Y^{MN}$ & $ \longrightarrow $ & $ (Y^{i0}, Y_{i0}) $ & $ \longrightarrow  $ & $ y^i=Y^{i0}$ & with \eqref{sec_sol_2} and $n_\pm=n_{\rm R,L}$ \\ 
   \hline 
     ${\rm O}(4, 4+n)$ &$ \hookleftarrow$ & $ {\rm O}(3,3) $ & $ \hookleftarrow $ & $ {\rm GL}(3)$ &  $D=6$ bosonic string,\\
   &&&&					& for \eqref{sec_sol_1}: with $n_{\rm V}=n$ \\
   &&&&					& for \eqref{sec_sol_2}:  with $n_-$=$n$  \\
   &&&&					& upon adding fermions: $\fr14$ SUSY \\ 	
   \hline 
  ${\rm O}(8, n+1)$ &$ \hookleftarrow$ & $ {\rm O}(7,7) $ & $ \hookleftarrow $ & $ {\rm GL}(7)$ &  for \eqref{sec_sol_1}: 
  $D=10$ bosonic string,  \\ 
  &&&&					& with $n_{\rm V} =n-7$ for $n\geq 7$\\
   &&&&					& upon adding fermions: $\fr12$ SUSY\\[1ex]
 ${\rm O}(8, n+1)$ &$ \hookleftarrow$ & $ {\rm O}(n,n) $ & $ \hookleftarrow $ & $ {\rm GL}(n)$ &  for \eqref{sec_sol_1}: 
  bosonic sector of  \\ 
   &&&&					& $D=n+3$ sugra, for $n\leq 7$\\
 &&&&					& upon adding fermions: $\fr12$ SUSY\\[1ex]
   ${\rm O}(8, n+1)$ &$ \hookleftarrow$ & $ {\rm O}(3,3) $ & $ \hookleftarrow $ & $ {\rm GL}(3)$ &  for \eqref{sec_sol_2}:
   $D=6$,  \\ 
  &&&&					& bosonic sector of $\mc{N}=(2,0)$ sugra,   \\
  &&&&					& $n-2$ tensor multiplets\\ 
\hline 
\end{tabular}
\caption{Table of gravitational theories which can be embedded into the present construction together with the corresponding solutions of the section constraint. Notations are the following: $n_{\rm s}$ --- number of scalar multiplets, $n_{\rm V}$ --- number of abelian vector multiplets, $n_{\pm}$ --- number of (anti)self-dual 2-forms}
\label{tab:sol}
\end{table}

\section{Generalized Scherk-Schwarz reduction}

In this section, we study reductions of the ${\rm O}(p,q)$ 
enhanced double field theory via a generalized Scherk-Schwarz
ansatz~\cite{Scherk:1979zr,Aldazabal:2011nj,
Geissbuhler:2011mx,Berman:2012uy,Musaev:2013rq,Aldazabal:2013mya,Lee:2014mla,Hohm:2014qga,Inverso:2017lrz}. 
We derive the consistency conditions on the Scherk-Schwarz twist matrices and rephrase them
as a generalized parallelizability condition. The particular structure of generalized diffeomorphisms (\ref{genLie})
and in particular the presence of constrained rotations in the diffeomorphism algebra requires a modification
of the standard constructions. 
We discuss in some detail the structure of three-dimensional gauge theories obtained by these 
generalized Scherk-Schwarz reductions.
We finally decompose the system of compatibility equations according to the solution
(\ref{sec_sol_1}) of the section constraints and reproduce as a particular case 
the structures known from ${\rm SL}(d+1)$ generalized geometry. In turn, this allows to employ known solutions
of this system in order to describe consistent truncations to three dimensions.

\subsection{Reduction ansatz and consistency equations}

The generalised Scherk-Schwarz reduction ansatz is encoded in an ${\rm O}(p,q)$ matrix $U_M{}^{\bN}(Y)$
and a weight factor $\rho(Y)$.
As in exceptional field theory~\cite{Hohm:2014qga}, 
we impose the following reduction ansatz on the fields
\bea
g_{\mu\nu}(x,Y) &=&\rho(Y)^{-2}\,{\rm g}_{\mu\nu}(x)\;,\nonumber\\
{\cal M}_{MN}(x,Y)&=&U_M{}^{\bM}(Y)\,U_N{}^{\bN}(Y)\,{\cal M}_{\bM\bN}(x)\;,\nonumber\\
{\cal A}_\mu{}^{MN}(x,Y)&=&\rho(Y)^{-1}\,U^M{}_{\bM}(Y)\,U^N{}_{\bN}(Y)\,A_\mu{}^{\bM\bN}(x)\;,\nonumber\\
{\cal B}_{\mu\,KL}(x,Y)&=&-\frac14\,\rho(Y)^{-1}\,
U^M{}_{\bN}(Y)\,\partial_{KL}U_{M\bM}(Y)\,A_\mu{}^{\bM\bN}(x)\;.
\label{SchSch}
\eea
Fundamental indices on the twist matrix are raised and lowered with the invariant tensor $\eta_{MN}$,
such that in particular $U_M{}^{\bM} U^{M\bN}=\eta^{\bM\bN}$\,.
Note that the ansatz for the constrained gauge connection ${\cal B}_{\mu\,KL}$ is manifestly
compatible with the constraints (\ref{sectionB}).
The gauge parameters $\Lambda^{MN}$, ${\Sigma}_{MN}$ associated with ${\cal A}_\mu{}^{MN}$, ${\cal B}_{\mu\,MN}$
factor accordingly
\bea
\Lambda{}^{MN}(x,Y)&=&\rho(Y)^{-1}U^M{}_{\bM}(Y)U^N{}_{\bN}(Y)\Lambda{}^{\bM\bN}(x)\;,\nonumber\\
\Sigma_{KL}(x,Y)&=&-\frac14\,\rho(Y)^{-1}\,U_{M\bN}(Y)\,\partial_{KL}U^M{}_{\bM}(Y)\,\Lambda{}^{\bM\bN}(x)\;.
\eea
The consistency constraints on the twist matrix are straightforwardly
obtained by working out the gauge transformations of these objects.
E.g.\ we find that
\bea
\mL_{(\L,\S)} \, g_{\mu\nu} &=&
2\,\rho^{-2}\left(
\L^{\bK\bL}\,\theta_{\bK\bL}\,{\rm g}_{\mu\nu} \right)
\;,\nonumber\\
\mL_{(\L,\S)}\,  {\cal M}_{MN} &=&
-2\,U_M{}^{\bM} U_N{}^{\bN}\left(
\Lambda^{\bK\bL}\,X_{\bK\bL,(\bM}{}^{\bQ}\,{\cal M}_{\bN)\bQ}
\right)
\;,
\label{actionM}
\eea
where the embedding tensor  $X_{\bK\bL,\bM}{}^{\bN}$  captures the gauge structure of
the three-dimensional theory, and is given by
\bea
\label{cons}
X_{\bK\bL,\bP\bQ}&=&\theta_{\bK\bL\bP\bQ}+
\frac12\,\left(
\h_{\bP[\bK} \theta_{\bL]\bQ}-
\h_{\bQ[\bK} \theta_{\bL]\bP}
\right)
+\theta\,\h_{\bP[\bK}\h_{\bL]\bQ}
\;,
\label{XTheta}
\eea
with the various components defined in terms of the twist matrix as
\bea
\theta_{\bK\bL\bP\bQ}&=&6\,\r^{-1}\,\dt_{LP}U_{N\,[\bK}U^N{}_{\bL}
U^L{}_{\bP}U^P{}_{\bQ]}\,
\;,\nonumber\\
\theta_{\bP\bQ}&=&4\,\r^{-1}\,U^K{}_{\bP}\,\dt_{KL}U^L{}_{\bQ}-\fr{4\,\r^{-1}}{p+q}\,\h_{\bP\bQ}U^{K\bL}\dt_{KL}U^L{}_{\bL}
-4\r^{-2}\,\dt_{\bP\bQ}\r\;,\nonumber\\
\theta&=&\fr{4\,\r^{-1}}{p+q}\,U^{K\bL}\dt_{KL}U^L{}_{\bL}\;.
\label{theta_comp}
\eea
The truncation (\ref{SchSch}) thus is consistent, if all the components (\ref{theta_comp})
of the embedding tensor are constant, i.e.\ 
\bea
\partial_{\bM} \theta_{\bK\bL\bP\bQ}&=& 0~=~
\partial_{\bM} \theta_{\bK\bL}~=~ \partial_{\bM} \theta
\;.
\label{consistency}
\eea
This provides a set of differential equations on the twist matrix and the weight factor
which encodes the consistency of the truncation.
In terms of ${\rm O}(p,q)$ representations, the
components (\ref{theta_comp}) of the embedding tensor transform as
\Yboxdim{0.9em}
\newcommand\sh{1.5}
\bea
 \Yvcentermath1 \gyoung(;,;) \;\otimes\; \gyoung(;,;) &\longrightarrow&
 \Yvcentermath1  
 \bullet  \;\oplus \;  \gyoung(;,;) \;\oplus \; \gyoung(;;) \;\oplus\;
 \gyoung(;,;,;,;)
\;\;\;,
\label{repsCC}
\eea
in a subrepresentation of the full tensor product (\ref{tensprod}).

For those theories admitting a supersymmetric embedding (i.e.\
the ${\rm O}(p,q)$ enhanced double field theories with $p=2,4,8$),
the structure (\ref{actionM}), (\ref{XTheta}) precisely reproduces
 the gauge structure of the associated three-dimensional gauged 
 supergravities~\cite{deWit:2003ja}.
 Here, that same structure appears more generally for an arbitrary group ${\rm O}(p,q)$.
The anti-symmetric tensor $\theta_{[\bP\bQ]}$ triggers three-dimensional gaugings 
in which the trombone scaling symmetry is part of the gauge group \cite{LeDiffon:2008sh}.
This follows directly from the first line of
(\ref{actionM}): a non-vanishining $\theta_{[\bP\bQ]}$ 
implies that the three-dimensional metric $g_{\mu\nu}$ is charged under part of the
gauge group.
The resulting theories do not admit a three-dimensional action and are defined
only on the level of the field equations. For most of the following discussions
we will thus require that $\theta_{[\bP\bQ]}=0$.

In a generic three-dimensional gauge theory, the embedding tensor (\ref{XTheta}) is subject to
the quadratic constraints
\bea
X_{\bK\bL\bP}{}^{\bR} X_{\bM\bN\bR}{}^{\bQ} -
X_{\bM\bN\bP}{}^{\bR} X_{\bK\bL\bR}{}^{\bQ} 
&=&
2\,X_{\bK\bL[\bM}{}^{\bR} X_{\bN]\bR\bP}{}^{\bQ} 
\;,
\eea
which guarantees closure of the gauge algebra. With the embedding tensor defined
by a twist matrix as (\ref{theta_comp}), these constraints follow directly from the section constraint (\ref{section}). 
Note that the section constraint combined with (\ref{theta_comp})
furthermore implies that
\bea
\theta_{[\bN_1 \dots \bN_4} \theta_{\bN_5 \dots \bN_8]} &=& 0
\;.
\label{Qext}
\eea
I.e.\ the generalized Scherk-Schwarz ansatz with twist matrices that obey the section condition
can only reproduce gaugings whose embedding tensor
satisfies the additional quadratic condition (\ref{Qext}). This is consistent with the fact, that the
general potential of $D=3$ half-maximal supergravity carries a term proportional to 
$\theta_{\bN_1 \dots \bN_4} \theta_{\bN_5 \dots \bN_8}\,{\cal M}^{\bN_1 \dots  \bN8}$,
with a scalar dependent totally antisymmetric tensor ${\cal M}^{\bN_1 \dots  \bN_8}$ \cite{Schon:2006kz},
that is not reproduced by the Scherk-Schwarz ansatz from the scalar potential given in (\ref{full_potential}).

\subsection{Generalized parallelizability}

Here we discuss the notion of \textit{generalized parallelizability} outlined in the introduction, 
which gives a more `geometric' perspective on the consistency conditions on the twist matrices 
discussed above. We claim that for the doubled tensor (in the sense of (\ref{Upsilon}))
 \be
  \mathfrak{U}_{\bar{M}\bar{N}} 
  \ \equiv \ \big(\rho^{-1} U^K{}_{[\bar{M}}{} U^L{}_{\bar{N}]}{}\;, \;-\frac{1}{4}\,\rho^{-1}(\partial_{KL} U^{P}{}_{\bar{M}}) U_{P\bar{N}}\big)\;, 
 \ee
which is manifestly compatible with the constraints on the second component by having the indices 
$KL$ be carried by a derivative, the consistency conditions can be stated simply in terms of 
the (generalized) Dorfman product  (\ref{DorfmanNN}) 
as 
 \be\label{genPARR}
  \mathfrak{U}_{\bar{M}\bar{N}} \ \circ \ \mathfrak{U}_{\bar{K}\bar{L}} \ = \
   -X_{\bar{M}\bar{N},\bar{K}\bar{L}}{}^{\bar{P}\bar{Q}}\, \mathfrak{U}_{\bar{P}\bar{Q}}\;. 
 \ee  
Here $X$ is the constant embedding tensor.  

We will now show that for the gauge vectors and its associated gauge symmetries 
the consistency of the Scherk-Schwarz ansatz is an immediate consequence of the fact that 
all relations are governed by the same Dorfman product `$\circ$' satisfying 
the Leibniz identity (\ref{LEIBNIZ}). 
We make the following Scherk-Schwarz ansatz      
for gauge fields and parameters:     
\be
\begin{split}
 \mathfrak{A}_{\mu}(x,Y) \ &= \ \mathfrak{U}_{\bar{M}\bar{N}}(Y)\,A_{\mu}{}^{\bar{M}\bar{N}}(x)\;, \\
 \Upsilon(x,Y) \ &= \ \mathfrak{U}_{\bar{M}\bar{N}}(Y)\,\Lambda^{\bar{M}\bar{N}}(x)\;. 
\end{split} 
\ee
It immediately follows with (\ref{DoubledVEctor}) and (\ref{SchSch}) that this is equivalent to 
the Scherk-Schwarz ansatz given above for the vector components. 
Let us now consider the gauge transformation of the Scherk-Schwarz ansatz: 
 \be
  \begin{split}
   \delta_{\Upsilon}\mathfrak{A}_{\mu}(x,Y) \ &= \ \partial_{\mu}\Upsilon -\mathfrak{A}_{\mu}\circ \Upsilon\\
   \ &= \ \mathfrak{U}_{\bar{M}\bar{N}}\,\partial_{\mu}\Lambda^{\bar{M}\bar{N}}
   -\mathfrak{U}_{\bar{K}\bar{L}}\circ \mathfrak{U}_{\bar{P}\bar{Q}}\, A_{\mu}{}^{\bar{K}\bar{L}}
   \Lambda^{\bar{P}\bar{Q}}\\
   \ &= \ \mathfrak{U}_{\bar{M}\bar{N}}\big(\partial_{\mu}\Lambda^{\bar{M}\bar{N}}
   +X_{\bar{K}\bar{L},\bar{P}\bar{Q}}{}^{\bar{M}\bar{N}} A_{\mu}{}^{\bar{K}\bar{L}}\Lambda^{\bar{P}\bar{Q}}\big)\\
   \ &= \ \mathfrak{U}_{\bar{M}\bar{N}}\,\delta_{\Lambda}A_{\mu}{}^{\bar{M}\bar{N}}\;, 
  \end{split}
 \ee  
where we used (\ref{genPARR}) and defined in the last line 
 \be
  \delta_{\Lambda}A_{\mu}{}^{\bar{M}\bar{N}} \ = \  \partial_{\mu}\Lambda^{\bar{M}\bar{N}}
   +X_{\bar{K}\bar{L},\bar{P}\bar{Q}}{}^{\bar{M}\bar{N}} A_{\mu}{}^{\bar{K}\bar{L}}\Lambda^{\bar{P}\bar{Q}}\;. 
 \ee
In here the $Y$-dependence encoded in $\mathfrak{U}(Y)$ has factored out, and this is  
precisely  the expected gauge transformation in gauged supergravity.    
Thus, the gauge transformations reduce consistently under Scherk-Schwarz. 
Similarly, one may show for all objects defined in terms of the Dorfman product, such as 
the non-abelian field strengths (\ref{frakFfieldstrength}), that they reduce consistently 
under Scherk-Schwarz. In general, the consistency conditions on the twist matrix are 
fully encoded in the algebra property  (\ref{genPARR}).

\subsection{\texorpdfstring{${\rm GL}(d+1)$}{GL(d+1)} twist equations }

In the following, we will be interested in constructing explicit solutions to the consistency
equations (\ref{consistency}). 
Obviously, the precise content of these equations will depend on the solution of the section 
constraints (\ref{section}), i.e.\ on the choice of physical coordinates
among the $\{Y^M\}$\,. We have discussed the different choices in subsection~\ref{subsec:solutions} above.
{
Let us stress that in this paper we will only be interested in constructing twist matrices that 
satisfy the section conditions (\ref{section}), i.e.\ in constructing consistent truncations from actual
higher-dimensional supergravities. It is known 
\cite{Berman:2012uy,Aldazabal:2013mya,Hohm:2014qga} that the 
match with lower-dimensional gauged supergravity 
formally 
holds even in the case the section
constraint is replaced by the weaker quadratic constraint on the resulting embedding tensor
(provided the initial scalar potential is written in an appropriate form).
On the other hand the higher-dimensional origin of the resulting gaugings within 
a well-defined theory remains mysterious.
}

As an ansatz for the solutions constructed in this section, we consider Scherk-Schwarz twist matrices $U_M{}^{\bN}(Y)$
living in the maximal ${\rm GL}(d+1)$ subgroup of ${\rm O}(d+1,d+1)$, i.e.\ of the explicit type
\begin{equation}
U_M{}^{\bM}=
\left[\begin{array}{c:cc}
\varphi \, V_A{}^{\bA}{}\ph{\Big[} & 0 \\ 
 \hdashline
0 & \varphi^{-1}\, (V^{-1})_{\bA}{}^A\ph{\Big[}
\end{array}\right]
\;,
\label{USL}
\end{equation}
with an ${\rm SL}(d+1)$ matrix $V_A{}^{\bA}$ and a scalar function $\varphi$.
Under this subgroup, the extended coordinates decompose as
\Yboxdim{0.9em}
\begin{equation}
\begin{aligned}
\left\{Y^{MN}\right\} & ~\longrightarrow~ \left\{ Y^{AB}, \; Y_{AB},\; Y_{A}{}^B\right\}\;,
\end{aligned}
\label{breakY}
\end{equation}
with the indices $A,B=1,\ldots,d+1$ labelling the fundamental representation of ${\rm SL}(d+1)$.
We moreover restrict the physical coordinates to $\left\{ Y^{AB}\right\}$, suppressing all dependence on \\
$\left\{  Y_{AB},Y_{A}{}^B\right\}$. This restriction is compatible with the choice (\ref{sec_sol_1})
of physical coordinates. What we will show in the following is that with this ansatz the 
consistency equations (\ref{theta_comp})--(\ref{consistency}) can be reduced 
to the ${\rm SL}(d+1)$ system
of equations that has been solved in~\cite{Hohm:2014qga} with solutions corresponding
to sphere and hyperboloid geometries.

Rather than directly plugging the ansatz (\ref{USL}) into the consistency equations (\ref{theta_comp})--(\ref{consistency}), it is useful to first analyze the representation content
of the latter. 
With the ansatz (\ref{USL}), the consistency equations (\ref{theta_comp}) 
turn into equations linear in the currents
\bea
J_{\bA\bB,\bC}{}^{\bD} &\equiv& 
(V^{-1})_{\bA}{}^A(V^{-1})_{\bB}{}^B\,
(V^{-1})_{\bC}{}^C\,\partial_{AB} V_{C}{}^{\bD} \;,\nonumber\\
j_{\bA\bB} &\equiv& \varphi^{-1}
\,(V^{-1})_{\bA}{}^A(V^{-1})_{\bB}{}^B\,\partial_{AB} \varphi
\;,
\eea
which under ${\rm SL}(d+1)$ transform in the representations
\bea{}
J_{\bA\bB,\bC}{}^{\bD} &:& 
[0,1,0,\dots,0] ~\oplus~ [2,0,0,\dots,0] 
~\oplus~ [0,0,1,\dots,0,1] ~\oplus~ [1,1,0,\dots,0,1]
\;, 
\nonumber\\
j_{\bA\bB} &:& [0,1,0,\dots,0]\;,
\label{repsJJ}
\eea
denoted by their standard Dynkin labels.
We may trace back the appearance of the various
components of these currents 
within the various components of the consistency equations by
decomposing the ${\rm O}(d+1,d+1)$ representations (\ref{repsCC}) of the latter 
under ${\rm SL}(d+1)$. Specifically, we find that the different components of the 
consistency equations (\ref{theta_comp})--(\ref{consistency})
accommodate the following components of the currents (\ref{repsJJ})
\Yboxdim{0.9em}
\bea
 \Yvcentermath1 
 \bullet &\longrightarrow&  - 
 \nonumber\\
 \Yvcentermath1   \gyoung(;,;) &\longrightarrow&  [0,1,0,\dots]
 \nonumber\\
   \gyoung(;;) &\longrightarrow&  [2,0,0,\dots]
 \nonumber\\
 \Yvcentermath1  \gyoung(;,;,;,;) &\longrightarrow&  [0,1,0,\dots] + [0,0,1,0,\dots,0,1]
\;\;\;.
\eea
We thus conclude that the consistency equations (\ref{theta_comp})--(\ref{consistency})
translate into
\bea
J_{\bA\bB,\bC}{}^{\bD} \Big|_{ [2,0,0,\dots,0] \;\oplus\; [0,0,1,\dots,0,1]} &=& {\rm const.}\;,
\label{con1}
\eea
together with two equations combining $j_{\bA\bB}$ with the projection $J_{\bA\bB,\bC}{}^{\bD} |_{ [0,1,0,\dots,0]}$ which take the explicit form
\bea
-\r^{-1}\varphi^{-2}
\Big(
\dt_{AB}(V^{-1})_{\bA\bB}{}^{AB} 
+(d-1)\, (V^{-1})_{\bA\bB}{}^{AB}{}
\,
 \dt_{AB}\,{\rm ln}\, \varphi
\Big)
&=&\theta_{\bA\bB\bC}{}^{\bC}~\stackrel!{=}~{\rm const}
\;,\nonumber\\
2\,\r^{-1}\varphi^{-2}\left(
\dt_{AB}(V^{-1})_{\bA\bB}{}^{AB} 
- 2\,(V^{-1})_{\bA\bB}{}^{AB}{}
\,
 \dt_{AB}\,{\rm ln}\, (\varphi \r)
\right)
&=& \theta_{[\bA\bB]}~\stackrel!{=}~{\rm const}
\;.\qquad
\label{con2}
\eea
It follows that with the ansatz
\bea
\rho&=&\varphi^{-(d+1)/2}
\;,
\eea
for the weight factor $\rho$,
these two equations coincide and the full system (\ref{con1})--(\ref{con2})
of consistency equations reproduces the ${\rm SL}(d+1)$ consistency
equations solved in~\cite{Hohm:2014qga} for sphere and hyperboloid compactifications.
In particular, for these solutions $\theta_{[\bA\bB]}=0=\theta_{\bA\bB\bC}{}^{\bC}$\,.
Translating the solutions of \cite{Hohm:2014qga} into our conventions here, we identify physical 
coordinates $\{y^i\}$, $i=1, \dots, d$, as (\ref{sec_sol_1}) among the $Y^{AB}$ and 
accordingly split the upper left block of (\ref{USL}) as
\bea
\varphi V_A{}^{\bA} &=&
\begin{pmatrix}
\varphi  V_0{}^0 & \varphi V_0{}^j\\
\varphi  V_i{}^0 & \varphi V_i{}^j
\end{pmatrix}
\nonumber\\
&=&
\left(\begin{array}{c:c}
(1-u)^{-1}\,(1+u\,k(u)) & -y^j\,(1-u)^{-1/2}\,k(u)\\  \hdashline
-y^i\,(1-u)^{-1/2} & \delta_i{}^{j}
\end{array}\right)
\;,
\label{Vexp}
\eea
with $u\equiv y^iy^i$, and with a scalar function $k(u)$ 
found as a solution of the differential equation 
\bea
2\,u\,(1-u)\,k'(u)&=&
\left( (d-1)\,u -d  \right) k(u) -1
\;.
\label{K}
\eea
The weight factor $\rho$ is given by\footnote{
To avoid confusion let us point out that $\rho$ in (\ref{rho}) denotes the weight factor 
of the ${\rm O}(d+1,d+1)$ consistency equations (\ref{theta_comp}) and {\em not} the
weight factor of the ${\rm SL}(d+1)$ equations in~\cite{Hohm:2014qga} from which 
it differs by a power of $\frac{d+1}{d-3}$\,. In particular, in the present context, the construction
applies to any values of $d$ without  an analogue of the relation (4.28) in \cite{Hohm:2014qga}.
This is due to the fact that the additional factor $\varphi$ in (\ref{USL}) has been fixed such
as to compensate for the missing powers of $(1-u)$.
}
\bea
\rho&=&(1-u)^{1/2}
\;.
\label{rho}
\eea
The resulting $U$-matrix (\ref{USL}) induces an embedding tensor 
$\theta_{(AB)}\propto\delta_{AB}$ in the
$\;\gyoung(;;)\;$ within (\ref{repsCC}). When evaluated in 
(\ref{actionM}), (\ref{cons}) it describes a gauge group 
\bea
{\rm G}_{\rm gauge} &=& {\rm SO}(d+1)\ltimes \TT^{d(d+1)/2}\;,
\label{gaugeG}
\eea
which is the semi-direct product of ${\rm SO}(d+1)$ with $\frac12\,d(d+1)$ nilpotent
generators transforming in the adjoint representation of ${\rm SO}(d+1)$.
It is important to note that the gauge sector of the 
resulting three-dimensional theory, obtained by
evaluating the action (\ref{action_fixed}) under the Scherk-Schwarz ansatz, 
is governed by a Chern-Simons action rather than a Yang-Mills action for the
vector fields. With the gauge group (\ref{gaugeG}) and the particular structure
of the embedding tensor (\ref{cons}), this theory may be rewritten as an
${\rm SO}(d+1)$ Yang-Mills gauge theory~\cite{Nicolai:2003bp} upon furthermore eliminating
$\frac12\,d(d+1)$ scalar fields from the action. The three-dimensional
scalar coset space then reduces from ${\rm SO}(d+1,d+1)/({\rm SO}(d+1)\times {\rm SO}(d+1))$
to ${\rm GL}(d+1)/{\rm SO}(d+1)$\,.
The generalized Scherk-Schwarz reduction in this case reproduces the 
consistent truncation of the $(d+3)$-dimensional bosonic string 
on the sphere $\SS^{d}$ which has been explicitly constructed in \cite{Cvetic:2000dm}.
In particular, it describes the $\SS^7$ reduction of the NS-NS sector of ten-dimensional
supergravity to an ${\cal N}=8$ half-maximal supergravity in three dimensions. The theory
does not admit an AdS$_3$ solution but a domain-wall solution that preserves half of the supersymmetry.

Note that here we have 
only given the explicit twist matrix for the case of compact gauge groups underlying sphere compactifications. It is straightforward to also employ the solutions from \cite{Hohm:2014qga} with non-compact gauge groups to describe consistent truncations on (warped)
hyperboloid backgrounds.

Let us finally stress that our construction of explicit twist matrices here has been based on 
restricting the coordinates to the antisymmetric bifundamental $\{Y^{AB}\}$ 
in the decomposition (\ref{breakY}) under ${\rm GL}(d+1)\subset{\rm O}(d+1,d+1)$\,.
In principle, one may also explore other choices of physical coordinates, e.g.\ within the 
adjoint representation $\{Y_A{}^{B}\}$ of ${\rm SL}(d+1)$, which together with an ansatz (\ref{USL})
for a ${\rm GL}(d+1)$ twist matrix will give rise to yet other solutions.

\section{Consistent truncations from \texorpdfstring{$D=6$}{D=6} dimensions}

In this section, we evaluate the generic reductions from the previous section
for the particular case of an $\SS^3$ reduction from $D=6$ dimensions.
As it turns out, in this case inequivalent reductions can be constructed
based on the alternative solution (\ref{sec_sol_2}) of the section constraint.
Moreover, the above constructed twist matrices admit a one-parameter deformation
corresponding to turning on an internal flux for the three-form field strength.
The resulting three-dimensional 
theories capture the compactification of six-dimensional
supergravities around the supersymmetric
AdS$_3\times \SS^3$ vacuum.

\subsection{Generic \texorpdfstring{$\SS^3$}{S3} reduction}

For $d=3$, the ${\rm GL}(4)$ twist matrix (\ref{USL}), (\ref{Vexp})
describes the generic $\SS^3$ reduction~\cite{Cvetic:2000dm}
from the minimal $D=6$, ${\cal N}=(1,0)$ supergravity coupled to a tensor multiplet. 
The total $D=6$ field content thus
combines the metric, a (non-chiral) two-form and a scalar field.
After $\TT^3$ reduction, this theory gives rise to a $D=3$ theory
with scalar coset space ${\rm SO}(4,4)/({\rm SO}(4)\times {\rm SO}(4)$.
It induces an embedding tensor of the form
\bea
\theta_{\bA\bB}&=& 4\,\delta_{\bA\bB}
\qquad\Longrightarrow\qquad
\Theta_{\bA\bB,\bC\bD} ~=~ 2\,\delta_{\bC[\bA}\delta_{\bB]\bD}
\;.
\label{theta1}
\eea
As described above, the resulting three-dimensional
theory is a Chern-Simons gauge theory with gauge group ${\rm SO}(4)\ltimes \TT^6$,
which may be rewritten as a more standard ${\rm SO}(4)$ Yang-Mills gauge theory
upon eliminating the six nilpotent gauge fields together with six of the scalar 
fields~\cite{Nicolai:2003bp}. The scalar coset space then reduces to ${\rm GL}(4)/{\rm SO}(4)$
and can be parametrized in terms of a symmetric ${\rm GL}(4)$ matrix $T_{AB}$.
The theory has a runaway potential given by~\cite{Cvetic:2000dm}
\bea
V&=&
4\left({\rm Tr}\,(T^2)-\frac12({\rm Tr}\,T)^2)\right)
\;,
\eea
and no ground state.

Interestingly, this solution allows for an alternative presentation upon using the
dual coordinates (\ref{sec_sol_2}). Switching from (\ref{sec_sol_1}) to these coordinates
and changing the twist matrix (\ref{USL}) into 
\begin{equation}
U_M{}^{\bM}(\tilde{y})=
\left(\begin{array}{c:cc}
\varphi\, (V^{-1})_{\bA}{}^{A}\ph{\Big[}  & 0 \\ 
 \hdashline
0 &\varphi^{-1}\,V_A{}^{\bA}\ph{\Big[}
\end{array}\right)\;, 
\label{UB}
\end{equation}
with $V$ still given by (\ref{Vexp}), produces another solution to the consistency equations
(\ref{theta_comp})--(\ref{consistency}),
with an embedding tensor given by
\bea
\theta_{\bA\bB\bC}{}^{\bD} &=& \varepsilon_{\bA\bB\bC\bE}\,\delta^{\bE\bD}
\;.
\label{theta2}
\eea
This is a DFT analogue of the construction~used in \cite{Malek:2015hma} to relate consistent
ExFT truncations from IIA and IB supergravity by accompanying the change of coordinates
by the action of an outer automorphism $V\rightarrow (V^T)^{-1}$ on the ${\rm SL}(4)$ twist matrix.
Here, the resulting gaugings are equivalent as can be seen by comparing the representations of the
embedding tensors  (\ref{theta1}), (\ref{theta2}) within ${\rm SO}(4,4)$
\bea
\theta_{\bA\bB} &\subset&  \gyoung(;;) ~=~ {\bf 35}_v 
\;,\qquad
 \theta_{\bA\bB\bC}{}^{\bD} ~\subset~  \Yvcentermath1   \gyoung(;,;,;,;) ~=~ {\bf 35}_s \oplus {\bf 35}_c
 \;.
 \label{35}
\eea
The two embedding tensors (\ref{theta1}), (\ref{theta2}) then are related by
a triality flip ${\bf 35}_v \leftrightarrow {\bf 35}_c$, the two gaugings hence equivalent.
They both describe the $\SS^3$
reduction of minimal $D=6$, ${\cal N}=(1,0)$ supergravity coupled to a tensor multiplet.

\subsection{\texorpdfstring{$D=6$, ${\cal N}=(1,0)$}{D=6 N=(1,0)} on \texorpdfstring{${\rm AdS}_3 \times \SS^3$}{AdS3xS3}}

In the three-dimensional case, the generic $\SS^3$ reduction constructed in \cite{Cvetic:2000dm}  
can be modified by integrating out the two-form from the resulting three-dimensional theory 
which gives rise to an additional contribution to the scalar potential. 
In turn, the new potential then supports a stable 
supersymmetric AdS$_3$ solution \cite{Deger:2014ofa},
corresponding to the supersymmetric ${\rm AdS}_3 \times \SS^3$ solution of
minimal $D=6$ supergravity.
For the description in terms of a Scherk-Schwarz twist matrix this corresponds to a deformation of the above construction by 
an extra matrix factor
\bea
{\cal U}(y) &=& U(y) \, \mathring{U}_\alpha(y)
\;,
\label{UU}
\eea
with the ${\rm GL}(4)$ matrix $U(y)$ from (\ref{USL}), (\ref{Vexp}), and the matrix $\mathring{U}(y)$ obtained by exponentiating 
some nilpotent generators of ${\rm SO}(4,4)$ according to
\bea
\mathring{U}_\alpha &=& {\rm exp}\left(\alpha\,(1+k(u))(1-u)^{-1/2} N_0\right)
\;,\nonumber\\
N_0&\equiv&
\begin{pmatrix}
0_{4\times 4}&n_0\\
0_{4\times 4}&0_{4\times 4}
\end{pmatrix}
\;,\qquad
n_0~\equiv~
\begin{pmatrix}
  0 &y^3&-y^2&0\\
 -y^3&0& y^1&0\\
  y^2&-y^1&0&0\\
0&0&0&0
\end{pmatrix}
\;,
\label{UBB}
\eea
with the function $k(u)$ from (\ref{K}) and a constant $\alpha$\,.
It is straightforward to check that the matrices $U$ and $\mathring{U}_\alpha$ commute and that their product (\ref{UU}) remains
a solution of the Scherk-Schwarz consistency equations. It results in an embedding tensor that in addition 
to (\ref{theta1}) has the further non-vanishing component
\bea
\theta_{\bA\bB\bC\bD} &=& -2\,\alpha\,\varepsilon_{\bA\bB\bC\bD}
\;.
\label{thetaA}
\eea
This gives rise to a three-dimensional gauging with the same gauge group ${\rm SO}(4)\ltimes T^6$
but a modified scalar potential 
\bea
V&=&
4\left({\rm Tr}\,(T^2)-\frac12\,({\rm Tr}\,T)^2+2\,\alpha^2\,{\rm det}\,T\right)
\;,
\label{VA}
\eea
which (for $\alpha=1$) exhibits a critical point at the scalar origin 
which corresponds to a supersymmetric AdS$_3$ solution~\cite{Deger:2014ofa}.
The product of twist matrices (\ref{UU}) thus describes the consistent truncation
of $D=6$, ${\cal N}=(1,0)$ supergravity on ${\rm AdS}_3 \times \SS^3$.

Similar to the discussion in the previous subsection, 
also the deformed twist matrix (\ref{UU}) can be expressed in
terms of the dual coordinates (\ref{sec_sol_2}).
In dual coordinates, the twist matrix $U$ in (\ref{UU}) is replaced by (\ref{UB})
whereas the factor $\mathring{U}_\alpha$ now is given by
\bea
\mathring{U}_\alpha(\tilde{y}) &=& {\rm exp}\left(\alpha\,(1+k(\tilde{u}))(1-\tilde{u})^{-1/2} N_0\right)
\;,\nonumber\\
N_0&\equiv&
\begin{pmatrix}
0_{4\times 4}&n_0\\
0_{4\times 4}&0_{4\times 4}
\end{pmatrix}
\;,\qquad
n_0~\equiv~
\begin{pmatrix}
  0 &0&0&\tilde{y}^1\\
  0 &0&0&\tilde{y}^2\\
  0 &0&0&\tilde{y}^3\\
0&-\tilde{y}^1&-\tilde{y}^2&-\tilde{y}^3
\end{pmatrix}
\;.
\eea
Again, the two matrices $U$ and $\mathring{U}_\alpha$ commute with their product solving
the Scherk-Schwarz conistency equations (\ref{theta_comp})--(\ref{consistency}).
The resulting embedding tensor turns out to be given by the sum of (\ref{theta2}) and (\ref{thetaA}):
\bea
\theta_{\bA\bB\bC}{}^{\bD} &=& \varepsilon_{\bA\bB\bC\bE}\,\delta^{\bE\bD}\;,\qquad
\theta_{\bA\bB\bC\bD} ~=~ -2\,\alpha\,\varepsilon_{\bA\bB\bC\bD}
\;,
\label{thetaAB}
\eea
inducing the same scalar potential (\ref{VA}). I.e.\ with respect to the decomposition
(\ref{35}) of the embedding tensor, its new component $\theta_{\bA\bB\bC\bD}$ lives in the ${\bf 35}_s$
and is not affected by the ${\rm SO}(4,4)$ triality flip ${\bf 35}_v \leftrightarrow {\bf 35}_c$\,.

\subsection{\texorpdfstring{${\cal N}=(1,1)$ and ${\cal N}=(2,0)$}{N=(1,1) and N=(2,0)} on \texorpdfstring{${\rm AdS}_3 \times \SS^3$}{AdS3xS3}}

We have presented the consistent truncations of $D=6$ ${\cal N}=(1,0)$ supergravity on $\SS^3$
described by an ${\rm SO}(4,4)$ twist matrix ${\cal U}$.
Upon embedding ${\rm SO}(4,4)$ into ${\rm SO}(4+m,4+n)$, the same twist matrix can be employed
to describe consistent truncation of $D=6$ supergravity coupled to vector or tensor multiplets.

E.g.\ choosing in the ${\rm SO}(8,4)$ theory physical coordinates according to (\ref{sec_sol_1}) 
together with a twist matrix (\ref{UU}) describes the consistent truncation of half-maximal 
$D=6$, ${\cal N}=(1,1)$ non-chiral supergravity on AdS$_3\times \SS^3$\,.
The embedding tensor of this theory is given by
the sum of (\ref{theta1}) and (\ref{thetaA}) as
\bea
\theta_{\bA\bB}&=& 4\,\delta_{\bA\bB}\;,\qquad
\theta_{\bA\bB\bC\bD} ~=~ -2\,\alpha\,\varepsilon_{\bA\bB\bC\bD}
\;,
\label{thetaN1}
\eea
for indices $\bA, \bB, \bC, \bD \in \{1,2,3,4\}$ and zero otherwise.
On the other hand, choosing for the ${\rm SO}(8,4)$ theory the 
dual physical coordinates according to (\ref{sec_sol_2}), together with a twist matrix (\ref{UB}), (\ref{UBB})
describes the consistent truncation of half-maximal 
$D=6$, ${\cal N}=(2,0)$ chiral supergravity (coupled to a tensor multiplet) on AdS$_3\times \SS^3$\,.
The embedding tensor of this theory is given by
(\ref{thetaAB}) as
\bea
\theta_{\bA\bB\bC}{}^{\bD} &=& \varepsilon_{\bA\bB\bC\bE}\,\delta^{\bE\bD}\;,\qquad
\theta_{\bA\bB\bC\bD} ~=~ -2\,\alpha\,\varepsilon_{\bA\bB\bC\bD}
\;,
\label{thetaN2}
\eea
for indices $\bA, \bB, \bC, \bD \in \{1,2,3,4\}$ and zero otherwise.
This is precisely the embedding tensor derived in \cite{Nicolai:2003ux} for the  
gauging associated with the ${\cal N}=(2,0)$ compactification (given in a different basis).
The present construction provides the full non-linear embedding of the three-dimensional theory
in six dimensions.
In this case, the gaugings induced by (\ref{thetaN1}) and by (\ref{thetaN2})
are no longer equivalent since the different representations (\ref{35})
of the embedding tensor are no longer related by triality within ${\rm SO}(8,4)$.
Accordingly, the higher-dimensional theories are strictly in-equivalent.
It is straightforward to extend the construction such as to include the couplings to
further ${\cal N}=(1,1)$ vector  or ${\cal N}=(2,0)$ tensor multiplets. The resulting
three-dimensional gaugings in particular reproduce the mass spectra 
computed in~\cite{Deger:1998nm,deBoer:1998kjm}.

\section{Conclusions and Outlook}

We have constructed enhanced double field theories in which the usual ${\rm O}(d,d)$ 
is enlarged to at least ${\rm O}(d+1,d+1)$ due to the inclusion of `dual graviton' graviton degrees of freedom. 
In this we have employed the `split formulation' common for exceptional field theory, in which one has 
external and internal coordinates. 
The structure of the resulting theory parallels maximal E$_{8(8)}$ ExFT \cite{Hohm:2014fxa}
and minimal SL(2) ExFT \cite{Hohm:2013jma}.
It can certainly be further generalized for other choices of groups together with
coordinates in the adjoint representation, c.f.\ the classifications in
\cite{Cremmer:1999du,deWit:1992up,deWit:2003ja,Strickland-Constable:2013xta}.
For three external dimensions the dual graviton components
arise among the `scalar' fields. One may also introduce the dual graviton in the more familiar 
`non-split' double field theory, for which they take the form of higher-rank ${\rm O}(d,d)$ representations, 
but so far this has only been achieved at the linearized level \cite{Bergshoeff:2016ncb,Bergshoeff:2016gub}. 
It remains as an open problem to find a non-split formulation for the dual graviton at the 
full non-linear level. 

The theories we have constructed for the groups ${\rm SO}(8,n)$ and ${\rm SO}(4,n)$ reproduce the bosonic
sectors of half-maximal and quarter-maximal supergravities, respectively. Depending on the solution of the 
section constraint, these theories describe chiral or non-chiral theories in six dimensions. It should be
straightforward and parallel to the maximal case \cite{Baguet:2016jph}
to introduce the fermion fields directly in the ExFT formulation given in this paper. This will require to
identify the proper ${\rm SO}(p)\times{\rm SO}(q)$ spin connections, determine their relevant components
via the torsionlessness condition (\ref{T0}) and work out the supersymmetric field equations.

As an application of these theories we have worked out a number of consistent truncations via the
generalized Scherk-Schwarz ansatz with suitably chosen twist matrices. In particular, the truncations
from six-dimensional supergravity on AdS$_3 \times \SS^3$ are constructed from a new class of twist matrices
that give rise to three-dimensional supergravities with supersymmetric ground states.
The consistent truncations of $D=6$, ${\cal N}=(1,1)$ and $D=6$, ${\cal N}=(2,0)$ supergravity on AdS$_3 \times \SS^3$
should be important in the context of 
the associated AdS/CFT dualities. It is interesting, that the reduction of the chiral ${\cal N}=(2,0)$ supergravity
appears consistent only in presence of an additional tensor multiplet which vanishes in the background.
It would be interesting to explore if similar consistent truncations can be constructed upon including massive
vector multiplets, leading to the three-dimensional gaugings constructed in \cite{Nicolai:2003ux}. The techniques recently developed in
\cite{Malek:2016bpu,Malek:2017njj} for generalized consistent truncations in exceptional field theory
may be very useful here.

We have found that the generalized Scherk-Schwarz ansatz cannot produce arbitrary three-dimensional gaugings
but only theories whose embedding tensor satisfies the additional condition (\ref{Qext}) --- at least as long as the
twist matrices satisfy the section constraints. A geometrical higher-dimensional origin of gaugings violating (\ref{Qext})
thus remains unclear. Similar no-go theorems have been found in \cite{Dibitetto:2012rk,Lee:2015xga,Malek:2015hma}
for higher-dimensional theories. Interestingly, most three-dimensional theories that seem to describe parts of the spectrum on
AdS$_3\times \SS^3\times \SS^3\times \SS^1$ appear to violate the condition (\ref{Qext})~\cite{Hohm:2005ui}. The
quest for consistent truncations around this background thus remains elusive. This may be related to recent
surprises in the BPS spectrum on this background~\cite{Eberhardt:2017fsi, Baggio:2017kza}. A notable exception is the lowest
massive spin-3/2 multiplet in the BPS spectrum which fits into a maximal three-dimensional supergravity~\cite{Hohm:2005ui}
whose ten-dimensional uplift may be constructible within maximal ExFT.

Another interesting generalization would be the explicit inclusion of Ramond-Ramond (RR) fields to the presented
formulation in order to enhance supersymmetry  from half-maximal to maximal. 
In the standard ${\rm O}(d,d)$ DFT the RR fields fit into spinor representations  
\cite{Hohm:2011zr,Hohm:2011dv} and it would be interesting to work out the generalization to 
the enhanced DFT discussed here. Since the extended section constraint allows for 
solutions corresponding to chiral supergravity in six dimensions it is tempting to speculate 
that the maximally supersymmetric and enhanced DFT may shed a light on 
Hull's conjectured six-dimensional  $(4,0)$ theory \cite{Hull:2000rr}.

\subsection*{Acknowledgements}
This work is partially supported by a PHC PROCOPE, projet No 37667ZL.
 The work of O.H. is supported by a DFG Heisenberg fellowship. The work of E.T.M. is supported by the Alexander von Humboldt Foundation and in part by the Russian Government programme of competitive growth of Kazan Federal University.

\appendix

\section*{Appendix}

\section{\texorpdfstring{${\rm O}(p,q)$}{O(p,q)} tensors and identities}

In this section, we present our ${\rm O}(p,q)$ conventions,
define a number of relevant tensors and collect some useful identities.
Generators $T_{MN}=T_{[MN]}$ of ${\rm O}(p,q)$ are labelled
by antisymmetric pairs of fundamental indices $M, N = 1, \dots, p+q$.
Their structure constants are given as
\bea
f_{PQ,MN}{}^{KL}&=&8\,\delta{}_{[P}{}^{[K} \eta_{Q][M} \delta{}_{N]}{}^{L]}
\;,
\eea
with the ${\rm O}(p,q)$ invariant tensor $\eta_{MN}$,
which we use to raise and lower indices.
The Cartan-Killing form is given by
\bea
\eta_{KL,MN} &\equiv& - \eta_{M[K}\eta_{L]N}\;.
\eea
The projector of a product of two adjoint representations onto the adjoint representation reads
\bea
\mathbb{P}^{PQ}{}_{RS}{}^{MN}{}_{KL} &=&
\frac1{16\,(p+q-2)}\,f^{UV,PQ}{}_{RS}{}f_{UV}{}^{MN}{}_{KL}
\nonumber\\
&=&
 \frac{1}{p+q-2}\,
\Big(
\delta_{[R}{}^{P} \delta_{S]}{}^{[M} \delta^{N]}{}_{[K} \delta_{L]}{}^Q 
-\delta_{R}{}^{[P} \eta^{Q][M}  \delta^{N]}{}_{[K} \eta_{L]S} 
\nonumber\\
&&{}
\qquad\qquad
-\delta_{[R}{}^{Q} \delta_{S]}{}^{[M} \delta^{N]}{}_{[K} \delta_{L]}{}^P 
+\delta_{S}{}^{[P} \eta^{Q][M}  \delta^{N]}{}_{[K} \eta_{L]R} 
\Big)
\;.
\label{ProjAdj}
\eea
We also define the tensor
\bea
s^{PQ,MN}{}_{KL}&=&8\,\delta_{(K}{}^{[P} \eta^{Q][M} \delta_{L)}{}^{N]}
\;,
\eea
symmetric under exchange of $[PQ]$ with $[MN]$, as well as the projector
\bea
\mathbb{A}^{PQMN}{}_{KLMN} &\equiv& \delta_{KLMN}{}^{PQMN}
\;.
\eea
In terms of these tensors, the ${\rm O}(p,q)$ section constraints (\ref{section}) can then be written as
\bea
\mathbb{A}^{PQMN}{}_{KLMN}\,\partial_{PQ}\otimes\partial_{MN} &=& 0 ~=~
\eta^{PM}\eta^{QN}\,\partial_{PQ}\otimes\partial_{MN}
\;,
\nonumber\\
s^{PQ,MN}{}_{UV}\,\partial_{PQ}\otimes\partial_{MN}&=& 0 
~=~f^{PQ,MN}{}_{UV}\,\partial_{PQ}\otimes\partial_{MN}
\;.
\label{sectionT}
\eea
A useful identity for the projection tensor (\ref{ProjAdj})
(the analogue of the ${\rm E}_{8(8)}$ identity (2.3) in \cite{Hohm:2014fxa})
is the following
\bea
\frac{p+q-2}{2}\,
\mathbb{P}^{PQ}{}_{RS}{}^{MN}{}_{KL}
&=&
-\frac14\left(\delta_{RS}{}^{PQ}\,\delta_{KL}{}^{MN}+\delta_{KL}{}^{PQ}\,\delta_{RS}{}^{MN}\right)
+\frac32\,{\mathbb A}^{PQMN}{}_{RSKL}{}
\nonumber\\
&&{}
+\frac1{64}\,s^{PQ,MN}{}_{UV} s_{RS,KL}{}^{UV}
+\frac1{64}\,f^{PQ,MN}{}_{UV} f_{RS,KL}{}^{UV}
\;,
\eea
which together with (\ref{sectionT}) shows in particular that
\bea
2\,(p+q-2)\,
\mathbb{P}^{PQ}{}_{RS}{}^{MN}{}_{KL}\,\partial_{PQ}\otimes\partial_{MN} 
&=&
-\left(\partial_{RS}\otimes\partial_{KL}  + \partial_{KL}\otimes\partial_{RS}  \right)
\;.
\eea
Another useful identity is given by
\bea
2\,(K-2)\,
\mathbb{P}^{PQ}{}_{RS}{}^{MN}{}_{KL}&=&
- f^{PQ,U[M}{}_{RS}\, \eta_{U[K} \delta^{N]}{}_{L]}
\;.
\eea

\section{\texorpdfstring{E$_{8(8)}$}{E8} generalized Dorfman structure}

For completeness we present in this appendix the generalized Dorfman product for E$_{8(8)}$, 
which allows one to formulate the gauge sector of the E$_{8(8)}$ ExFT constructed in \cite{Hohm:2014fxa} 
in the same way as in sec.~\ref{genDorfmanSEC}. 
We use the same notation and conventions as in \cite{Hohm:2014fxa}, to which we refer the reader for 
further details. In particular, $M,N,\ldots=1,\ldots, 248$ 
denote the adjoint ${\rm E}_{8(8)}$ index. 
We group the two gauge parameters, as in the main text, into the `doubled' object 
 \be\label{doubledE8parameter}
  \Upsilon \ = \ \big(\Lambda^M, \Sigma_M\big)\;, 
 \ee
and assume that the second component is a covariantly constrained object.    
The generalized Lie 
derivative of an adjoint vector with density weight $\lambda$ can then be written as 
 \be\label{usualgenLie}
  \mathbb{L}^{[\lambda]}_{\Upsilon}V^M \ = \ \Lambda^N\partial_N V^M 
  +f^{M}{}_{NK} R^N({\Upsilon}) V^K +\lambda\, \partial_N\Lambda^N V^M\;, 
 \ee
where we defined
 \be\label{Rtensor}
  R^M(\Upsilon) \ \equiv \ f^{MN}{}_{K}\,\partial_N\Lambda^K +\Sigma^M\;. 
 \ee
We recall from \cite{Hohm:2014fxa} that $\Lambda^M$ has weight one, $\Sigma_M$ has weight zero, 
and $\partial_M$ lowers the weight by one, $[\partial_M]=-1$, so that 
$R^M$ has weight zero.  
The above Lie derivatives close according to the `E-bracket', $[ \mathbb{L}_{\Upsilon_1},\mathbb{L}_{\Upsilon_2}] =  \mathbb{L}_{[\Upsilon_1,\Upsilon_2]}$, whose explicit form we will give momentarily. 
A useful intermediate relation for proving closure, in terms of (\ref{Rtensor}), is  
 \be\label{MasterRRel}
  R_M([\Upsilon_1,\Upsilon_2]) \ = \ 2\, \Lambda_{[1}{}^{N}\partial_N R_M(\Upsilon_{2]}) 
  \ + \ f_{MNK} R^N(\Upsilon_1)R^K(\Upsilon_2)\;. 
 \ee
We also recall that there are trivial gauge parameters with respect to 
which the generalized Lie derivatives act trivially on fields as a consequence of the section constraints. 
They take the form 
 \be\label{trivPARA}
 \begin{split}
  &\Lambda^M \ = \ \eta^{MN}\Omega_N\;, \qquad \text{with $\Omega_M$ covariantly constrained}\;, \\
  &\Lambda^M \ = \ (\mathbb{P}_{3875})^{MK}{}_{NL}\, \partial_K\chi^{NL}\\
  & \Lambda^M \ = \ f^{MN}{}_{K}\,\Omega_{N}{}^{K}\;, \qquad 
  \Sigma_M \ = \ \partial_M\Omega_{N}{}^{N}+\partial_N\Omega_M{}^{N}\;, 
 \end{split}
 \ee
where $\Omega_M{}^{N}$ is covariantly constrained in the first index.

Let us now turn to the definition of the generalized Dorfman product in terms of the doubled 
vectors (\ref{doubledE8parameter}):  
 \be\label{E8DORFMAN}
  \Upsilon_1 \circ  \Upsilon_2 \ \equiv \ \Big(\,\mathbb{L}_{\Upsilon_1}^{[1]}\Lambda_2{}^M \;,\; \,
  \mathbb{L}_{\Upsilon_1}^{[0]}\Sigma_{2M} \ + \ \Lambda_2{}^N\partial_M R_{N}(\Upsilon_1)\,\Big)\;. 
 \ee
This definition is such that  the E-bracket is given by 
 \be\label{E8Ebracket}
  \big[\Upsilon_1,\Upsilon_2\big] \ = \ \frac{1}{2}\big(\Upsilon_1 \circ  \Upsilon_2 
  -\Upsilon_2  \circ  \Upsilon_1 \big)\;. 
 \ee
More precisely, this agrees with the bracket given in \cite{Hohm:2014fxa} upon adding a trivial parameter 
of the last form in (\ref{trivPARA}), with $\Omega_N{}^K  =  \Sigma_{[1 N}\Lambda_{2]}{}^K$, which is manifestly compatible 
with the constraint.   
On the other hand, the symmetric part of the product is trivial: One finds by an explicit computation 
 \bea
   \frac{1}{2}\big(\Upsilon_1 \circ  \Upsilon_2 
   + \Upsilon_2  \circ  \Upsilon_1 \big)   & = & \Big(\,7\,(\mathbb{P}_{3875})^{MK}{}_{NL}\, 
   \partial_K\big(\Lambda_1^N\Lambda_2^L\big) + \frac{1}{8}\, \partial^M\big(\Lambda_1^N \Lambda_{2N}\big)
   +f^{MN}{}_{K}\, \Omega_{N}{}^{K}\;, \nonumber\\
   &&\qquad \partial_M\Omega_N{}^{N} + \partial_N\Omega_{M}{}^{N}\;\Big)\;, 
\eea
where 
 \be
  \Omega_{M}{}^{N} \ \equiv \ \Lambda_{(1}{}^{N}\Sigma_{2)M}
  -\frac{1}{2}\, f^{N}{}_{KL}\,\Lambda_{(1}{}^{K}\,\partial_M\Lambda_{2)}{}^L\;. 
 \ee
Since $\Omega_{M}{}^{N}$ so defined is manifestly covariantly constrained in the  first index, 
this is indeed a trivial parameters  of the last form in (\ref{trivPARA}). 

We next prove that the Dorfman product satisfies the Leibniz algebra relation discussed in the main text. 
To this end we define again an extended generalized Lie derivative on 
doubled vectors $\mathfrak{A}=(A^M, B_M)$ according to 
 \be
  \mathbb{L}_{\Upsilon}{\mathfrak A} \ \equiv \ \Upsilon \circ {\mathfrak A}\;, 
 \ee
and verify that they satisfy the same algebra w.r.t.~(\ref{E8Ebracket}):  
 \be\label{MasterCLosure}
  \big[ \mathbb{L}_{\Upsilon_1},\mathbb{L}_{\Upsilon_2}\big]{\mathfrak A} \ = \ \mathbb{L}_{[\Upsilon_1,\Upsilon_2]}{\mathfrak A}\;. 
 \ee
This relation only needs to be proved when acting on 
the second, covariantly constrained component of $\mathfrak{A}$, 
for which closure  can be quickly seen to be equivalent to 
 \be\label{closureEXT}
  \partial_M R_N([\Upsilon_1,\Upsilon_2]) \ = \ 
  \mathbb{L}_{\Upsilon_1}^{[-1]}\big(\partial_MR_N(\Upsilon_2)\big)    
  - \mathbb{L}_{\Upsilon_2}^{[-1]}\big(\partial_MR_N(\Upsilon_1)\big)    \;. 
 \ee   
This in turn can be proved by 
taking  the derivative of (\ref{MasterRRel}) 
and using the Lemma   (2.13) of \cite{Hohm:2014fxa}. The proof of the Leibniz identity (\ref{LEIBNIZ}) 
finally follows precisely as in the main text.

Let us now turn to the definition of an invariant inner product on the space of doubled vectors in order 
to construct a Chern-Simons action. 
The following symmetric pairing transforms covariantly (i.e.~as a scalar density of weight one 
in the sense of (\ref{COVcondINNER}))
 \be
  \langle\langle \mathfrak{A}_1, \mathfrak{A}_{2}\rangle \rangle \ = \ 
  2\, A_{(1}{}^{M} B_{2)M} - f^{K}{}_{MN} A_{(1}{}^M \partial_K A_{2)}{}^N\;.  
 \ee
In order to prove this covariance property  one has to compute the non-covariant 
variation of the second term, which in turn cancels the effect of the `anomalous' 
term in the definition of the Dorfman product (\ref{E8DORFMAN}). 
Specifically, we have to establish  
 \be
 \begin{split}
  \Delta_{\Upsilon}\big(   f^{K}{}_{MN} A_{(1}{}^M \partial_K A_{2)}{}^N\big) \ &= \ 
  f^{M}{}_{NK} f^{K}{}_{PQ} A_{(1}{}^{N} \partial_M R^P(\Upsilon) A_{2)}{}^{Q}\\
  \ &= \ 2\,A_{(2}{}^M A_{1)}{}^N \partial_M R_{N}(\Upsilon)
  \;,  
 \end{split}
 \ee
which follows by a somewhat tedious computation, writing out $R$ and using Lemma (2.13) and (A.1) in \cite{Hohm:2014fxa} in order to reduce the 
number of $f$'s.
Given the covariance property, it follows that under an integral we have an invariant inner product: 
 \be
  \langle  \mathfrak{A}_1, \mathfrak{A}_{2}\rangle \ \equiv \ \int {\rm d}^{248}Y\big(
  A_1{}^M B_{2M}+A_2{}^M B_{1M} - f^{M}{}_{NK}A_1{}^{N} \partial_MA_2{}^{K}\big)\;, 
 \ee 
where the second term was simplified by integration by parts. 
We can rewrite this as 
 \be
   \langle  \mathfrak{A}_1, \mathfrak{A}_{2}\rangle \ \equiv \ \int {\rm d}^{248}Y\big(
   A_1{}^{M}R_{M}(\mathfrak{A}_2) +A_2{}^M B_{1M}\big)\;. 
 \ee
This form makes it manifest, as in the main text, that if one argument is trivial the inner product is zero, 
c.f.~(\ref{degeneracy}). 

With the above we have established that the analogues of all Dorfman-type 
identities used in the main text also hold for the ${\rm E}_{8(8)}$ case. 
This implies that the discussion of covariant derivatives, gauge fields and the tensor hierarchy 
proceeds in complete parallel. In particular, there is a (generalized) Chern-Simons formulation 
for the (doubled) gauge vector $\mathfrak{A}_{\mu}$ for the ${\rm E}_{8(8)}$ ExFT 
that takes precisely the same form as (\ref{SCS}).


\providecommand{\href}[2]{#2}\begingroup\raggedright\endgroup

\end{document}